\documentclass[aps,prd,amsmath,amssymb,preprintnumbers,nofootinbib,a4paper,11pt]{revtex4-1}
\pdfoutput=1
\usepackage{amsthm}
\usepackage{graphicx}
	\graphicspath{{./figures/}}
	
\usepackage{url}
\usepackage{youngtab}
\usepackage[bookmarks, pagebackref=false]{hyperref}
\usepackage{color}
	\definecolor{rossoCP3}{cmyk}{0,.88,.77,.40}
		\definecolor{graa}{rgb}{0.8,0.8,0.8}
		\definecolor{blaa}{rgb}{0.2,0.2,0.6}
		\hypersetup{
			colorlinks, 
			bookmarksopen, 
			bookmarksnumbered,
			citecolor=blaa, 		
			linkcolor=rossoCP3,	
			urlcolor=rossoCP3,			
			}
\usepackage{dcolumn}
\usepackage{bm}
\usepackage{bbm}
\usepackage{pxfonts}

\usepackage{placeins}
\usepackage{xspace}
\usepackage{cancel} 

\usepackage{slashed}
\usepackage[caption=false, labelformat=simple, listofformat=subsimple, labelfont=bf, margin=5pt, justification=raggedright]{subfig}

	\makeatletter
		\renewcommand{\p@subfigure}{}
	\makeatother

\newcommand{\ea}[1]{
\begin{align}
#1
\end{align}
}

\newcommand{\beq}{\begin{eqnarray}}
\newcommand{\eeq}{\end{eqnarray}}

\newcommand{\bmp}{\noindent\begin{minipage}{16cm}}
\newcommand{\emp}{\end{minipage}\vskip 7mm} 

\newcommand{\eff}{\textrm{eff}}
\newcommand{\atheorem}{\emph{\lowercase{a}}~theorem\xspace}
\newcommand{\ctheorem}{\emph{\lowercase{c}}~theorem\xspace}
\newcommand{\afunction}{\emph{\lowercase{a}}~function\xspace}

\newcommand{\betafunction}{$\beta$~function\xspace}
\newcommand{\betafunctions}{$\beta$~functions\xspace}
\newcommand{\nonBZ}{{\cancel{BZ}}}
\newcommand{\afree}{a_\textrm{free}}

\newcommand{\yf}{\tiny \yng(1)}

\DeclareMathOperator{\tr}{Tr}

\def\lsim{\mathrel{\rlap{\lower4pt\hbox{\hskip1pt$\sim$}}
    \raise1pt\hbox{$<$}}}                
\def\gsim{\mathrel{\rlap{\lower4pt\hbox{\hskip1pt$\sim$}}
    \raise1pt\hbox{$>$}}}                

\begin{document}
\title{\texorpdfstring{\Large\color{rossoCP3} The \atheorem for Gauge--Yukawa theories beyond Banks--Zaks}{ The \atheorem for Gauge--Yukawa theories beyond Banks--Zaks}}
\author{Oleg Antipin}
\email{antipin@cp3.dias.sdu.dk} 
\author{Marc Gillioz}
\email{gillioz@cp3.dias.sdu.dk} 
\author{Esben M\o lgaard}
\email{molgaard@cp3.dias.sdu.dk} 
\author{Francesco Sannino}
\email{sannino@cp3.dias.sdu.dk}  
 \affiliation{
{ \color{rossoCP3}  \rm CP}$^{\color{rossoCP3} \bf 3}${\color{rossoCP3}\rm--Origins} \& the Danish Institute for Advanced Study {\color{rossoCP3} \rm DIAS},\\ 
University of Southern Denmark, Campusvej 55, DK--5230 Odense M, Denmark.
} 
\begin{abstract}
We investigate the \atheorem for nonsupersymmetric gauge--Yukawa theories beyond the leading order in perturbation theory. The exploration is first performed in a model-independent manner and then applied to a specific relevant example. Here, a rich fixed point structure appears including the presence of a merging phenomenon between non--trivial fixed points for which the \atheorem has not been tested so far. 

\vspace{0.5cm}
\noindent
{ \footnotesize  \it Preprint: CP$^3$-Origins-2013-3 \& DIAS-2013-3}
\end{abstract}

\maketitle

\section{Introduction}

The recent work by Komargodski and Schwimmer~\cite{Komargodski:2011vj, Komargodski:2011xv} on the \atheorem has attracted much interest. The \atheorem is thought to be the generalization to four dimensions of Zamolodchikov's \ctheorem~\cite{Zamolodchikov:1986gt} which establishes the irreversibility of the renormalization group (RG) flow in two dimensions. Several extensions of the work of Zamolodchikov to four dimensions have been proposed. In particular, Cardy pointed to the function $a$ given by the integral of the trace of the energy--momentum tensor over the four--sphere~\cite{Cardy:1988cwa}. He then conjectured that $a$ has properties similar to its two dimensional cousin $c$. Cardy's conjecture was tested within the framework of perturbation theory in~\cite{Osborn:1989td,Jack:1990eb,Osborn:1991gm}. It was discovered that the \afunction is not monotonically decreasing along the RG flow, and an alternative function, denoted here by $\tilde{a}$, was devised to be monotonically decreasing in perturbation theory. A nice property of $\tilde{a}$ is that it coincides with Cardy's \afunction at fixed points, where the \betafunctions vanish. 

Komargodski and Schwimmer suggested a non--perturbative proof of the \atheorem  by introducing a dilatonic field \cite{Komargodski:2011vj,Komargodski:2011xv} whose scattering amplitude is related to the \afunction. In fact, the analyticity of the amplitude supports the monotonicity of the $\tilde{a}$ function (not $a$) along the RG flow~\footnote{The relation between the trace anomaly and the amplitude of the three point function of the energy--momentum tensor was already pointed out in~\cite{Cappelli:2000dv, Cappelli:2001pz}.}. Establishing the existence of a function that is monotonic along the RG flow can lead to relevant constraints on the dynamics of a given gauge theory. In supersymmetry, the existing relation between $a$ and the $R$--charge~\cite{Anselmi:1997am, Anselmi:1997ys} found use in the $a$--maximization \cite{Intriligator:2003jj}. Using holographic methods, the \atheorem could be tested and verified in the context of supersymmetry~\cite{Freedman:1999gp, Girardello:1998pd}.  The generalization, using holography, to arbitrary space-time dimensions was explored in~\cite{Myers:2010tj}. More recently, the relation between scale and conformal invariance in 4D was elucidated in the context of the \atheorem~\cite{Fortin:2012hn, Luty:2012ww}.

The goal of this work is to extend the perturbative analyses of the \atheorem for nonsupersymmetric gauge theories with fermions and gauge singlet scalars, interacting via Yukawas, to the maximum known order in perturbation theory. This allows us to investigate the details of the \atheorem, particularly for non--standard fixed point structures.   

The paper is structured as follows. In Section~\ref{3loops} we introduce the essential tools and discuss the generic expression of the $\tilde{a}$ function relevant for its determination to three loops in the gauge coupling. Concrete examples are presented in Section \ref{example} where we explicitly evaluate $\tilde{a}$ at the Banks--Zaks (BZ) fixed point (FP) for a gauge--Yukawa theory to the leading and the next--to--leading order in perturbation theory.  We then discover, for a certain non asymptotically free gauge theory, the appearance of a perturbative UV fixed point (UVFP), and at the next--to--leading order also the appearance of another infrared (IR) fixed point. By changing the number of matter fields we observe the merging and disappearance of both fixed points. This rich structure of fixed points constitutes an interesting playground for elucidating the properties of the $\tilde{a}$ function in gauge theories. We conclude in Section~\ref{conclusions} and provide a number of technical details in Appendix~\ref{appendix}.

\section{The \atheorem beyond the leading order}
\label{3loops} 

In four dimensions and for a general quantum field theory the vacuum expectation value of the trace of the energy--momentum tensor for a locally flat metric $g_{\mu \nu}$ reads
\begin{equation}
	\left\langle T_\mu^\mu \right\rangle = c \, W^2(g_{\mu\nu})
		- a \, E_4(g_{\mu\nu}) + \ldots \ ,
\end{equation}
where $a$ and $c$ are real coefficients, $E_4(g_{\mu\nu})$ the Euler density and $ W(g_{\mu\nu})$ the Weyl tensor. The dots represent contributions coming from operators that can be constructed out of the fields defining the theory. Their contribution is proportional to the \betafunctions of their couplings. The coefficient $a$ is the one used in Cardy's conjecture, and for a free field theory it is~\cite{Duff:1977ay}
\begin{equation}
	\afree = \frac{1}{90 (8\pi)^2} \left( n_s + \frac{11}{2} n_f + 62 n_v \right) \ ,
	\label{eq:afree}
\end{equation}
where $n_s$, $n_f$ and $n_v$ are respectively the number of real scalars, Weyl fermions and gauge bosons. 

The change of $a$ along the RG flow is directly related to the underlying dynamics of the theory via the \betafunctions. This can be shown by exploiting the abelian nature of the trace anomaly which leads to the Weyl consistency conditions in much the same manner as the well known Wess--Zumino consistency conditions~\cite{Wess:1971yu}. As discussed in the introduction and following the work of Jack and Osborn~\cite{Osborn:1989td, Jack:1990eb},  rather than using $a$ one uses the function $\tilde{a}$ related to it by
\begin{equation}
	\tilde{a} = a + W_i \beta_i \ ,
	\label{eq:atilde}
\end{equation}
where $W_i$ is a one--form which depends on the couplings of the theory. The Weyl consistency conditions\footnote{It is worth mentioning that the Weyl consistency conditions used above assume that the trace of the energy--momentum tensor vanishes when all the \betafunctions are zero simultaneously. Exceptions are known to exist~\cite{Fortin:2012cq} and in this case one modifies the consistency conditions~\cite{Fortin:2012hn,Luty:2012ww} in order to build $\tilde{a}$.} imply for $\tilde{a}$ 
\begin{equation}
	\partial_i \tilde{a} = -\chi_{ij} \beta_j + (\partial_i W_j - \partial_j W_i) \beta_j \ ,
	\label{eq:consistencycondition}
\end{equation}
where $\chi_{ij}$ can be viewed as a metric in the space of couplings. The positivity of the metric $\chi$ is established in perturbation theory, and therefore in this regime the function $\tilde{a}$ is monotonic along the RG flow
\begin{equation}
	\mu \frac{d\tilde{a}}{d\mu} = \beta_i \partial_i \tilde{a}
		= -\chi_{ij} \beta_i \beta_j \leq 0 \ .
\end{equation}
The irreversibility of the RG flow has been conjectured to be valid beyond perturbation theory.  

In the next subsections, capitalizing on the work of Jack and Osborn \cite{Jack:1990eb}, we will construct $\tilde{a}$ for gauge theories featuring fermionic matter interacting with gauge singlet scalar fields via Yukawa interactions to the highest known order in perturbation theory.

\subsection{Gauge--Yukawa Theories}

We consider  the following Lagrangian skeleton 
\begin{equation}
	\mathcal{L} = -\frac{1}{4g^2} F_{\mu\nu} F^{\mu\nu}
		+ i \, \Psi_i^\dag \sigma^\mu D_\mu \Psi_i
		+ \frac{1}{2} \partial_\mu \phi_a \partial^\mu \phi_a
		- \left( y^a_{ij} \, \Psi^c_i \Psi_j \phi_a + \textrm{h.c.} \right)
		- \frac{1}{4!} \lambda_{abcd} \, \phi_a \phi_b \phi_c \phi_d \ ,
	\label{eq:Lagrangian}
\end{equation}
where we dropped the gauge indices for $F_{\mu\nu}$ and the Weyl fermions $\Psi_i$. The  fermions transform according to a given representation $R$ of the underlying gauge group. The real scalars $\phi$ are singlets with respect to the gauge group. The indices run over the number of matter fields. To exemplify our results we will consider gauge theories for which the Yukawa and quartic interactions depend each on a single parameter as follows
 \begin{equation}
	y^a_{ij} \equiv y \, T^a_{ij} \ ,
	\quad\quad
	\lambda_{abcd} \equiv \lambda \, T_{abcd} \ ,
	\label{eq:tensorcouplings}
\end{equation}
where the $T$ are tensors specified in a given theory. There are therefore three couplings in our setup: gauge $g$, Yukawa $y$ and quartic $\lambda$. We define
\begin{equation}
	\alpha_g = \frac{g^2}{(4 \pi)^2},
	\quad
	\alpha_y = \frac{y^2}{(4 \pi)^2},
	\quad
	\alpha_\lambda = \frac{\lambda}{(4 \pi)^2}.
	\label{eq:alphas}
\end{equation}
The generic structure of the associated \betafunctions ($\beta_{\alpha} = \partial \alpha/ \partial \ln \mu$) reads\footnote{The factors of 2 in the definition of the gauge and Yukawa \betafunctions follow from the definitions (\ref{eq:alphas}). One has for example $\beta_{\alpha_g} / \alpha_g = 2 \beta_g / g$.}
\begin{eqnarray}
	\beta_{\alpha_g} & = & -2\alpha_g^2 \left[b_0 + b_1 \alpha_g + b_y \alpha_y 
		+b_2 \alpha_g^2 + b_3 \alpha_g \alpha_y + b_4 \alpha_y^2\right] \ ,
	\label{eq:betagauge} \\
	\beta_{\alpha_y} & = & 2\alpha_y \left[c_1 \alpha_y + c_2 \alpha_g
		+ c_3 \alpha_g \alpha_y + c_4 \alpha_g^2 + c_5 \alpha_y^2
		+ c_6 \alpha_y \alpha_\lambda + c_7 \alpha_\lambda^2\right] \ ,
	\label{eq:betaYukawa} \\
	\beta_{\alpha_\lambda} & = & d_1 \alpha_\lambda^2+d_2 \alpha_\lambda \alpha_y +d_3 \alpha_y^2 \ .
	\label{eq:betaquartic}
\end{eqnarray}
The expansion to three loops in the gauge coupling, two loops in the Yukawa and one in the quartic coupling leads to a consistent expression for $\tilde{a}$ to order $\alpha^3$. If the scalars were charged under the gauge group, terms proportional to $\alpha_g \alpha_\lambda$ would appear.

Having established the generic form of the \betafunctions, we move to determining the metric $\chi$ and one--form $W$. They can be found by examining the relevant Feynman diagrams which enter the computation of the trace anomaly, as shown in Appendix~\ref{appendix}. We find
\begin{equation}
	\chi =
	\left(\begin{array}{ccc}
		\frac{\chi_{gg}}{\alpha_g^2} \left(1 + A \alpha_g + B_1 \alpha_g^2 + B_2 \alpha_g \alpha_y \right) & B_0 & 0 \\
		B_0 & \frac{\chi_{yy}}{\alpha_y} \left( 1 + B_3 \alpha_y + B_4 \alpha_g \right) & 0 \\
		0 & 0 & \chi_{\lambda\lambda}
	\end{array}\right) \ .
	\label{eq:metric}
\end{equation}
The coefficient $\chi_{gg}$ enters at the one--loop order, $A$ and $\chi_{yy}$ at two loops, while $\chi_{\lambda\lambda}$ and the $B_i$'s appear only at three loops. Similarly, the one--form $W$ takes the form
\begin{eqnarray}
	W_g & = & \frac{1}{\alpha_g} \left(D_0 + D_1 \alpha_g 
		+ C_1 \alpha_g^2 + C_2 \alpha_g \alpha_y \right) \ ,\nonumber \\
	W_y & = & D_2 + C_3 \alpha_y + C_4 \alpha_g \ ,  \label{eq:oneform} \\
	W_\lambda & = & D_3 \alpha_\lambda \nonumber \ .
\end{eqnarray}
The general structure of $\chi$ confirms that it is sufficient for all our purposes to consider the Yukawa \betafunction (\ref{eq:betaYukawa}) to two--loop order and the quartic one (\ref{eq:betaquartic}) to one--loop only.

The leading coefficients $\chi_{gg}$, $\chi_{yy}$ and $\chi_{\lambda\lambda}$ are~\cite{Jack:1990eb}
\begin{equation}
	\chi_{gg} = \frac{d(G)}{128 \pi^2},
	\quad\quad
	\chi_{yy} = \frac{1}{128 \pi^2}
		\left( \frac{1}{3} T^a_{ij} T^{a*}_{ij} \right),
	\quad\quad
	\chi_{\lambda\lambda} = \frac{1}{128 \pi^2}
		\left( \frac{1}{72} T_{abcd} T_{abcd} \right) \ ,
	\label{eq:chis}
\end{equation}
where we used the tensors $T$ defined in eq.~(\ref{eq:tensorcouplings}) and $d(G)$ denotes the dimension of the adjoint representation $G$ of the underlying gauge group, i.e.~the number of gluons. $A$ is given by~\cite{Jack:1990eb}
\begin{equation}
	A = 17 C_2(G) - \frac{10}{3} N_R \, T(R), 
	\label{eq:A}
\end{equation}
where $C_2(G)$ is the quadratic Casimir of the adjoint, $N_R$ the number of Weyl fermion in the representation $R$ and $T(R)$ is the trace normalization satisfying $T(R) \delta_{ab} = \tr(R_a R_b)$, $R_a$ being the generators of the fermions under the gauge group. With these coefficients, the metric $\chi$ is positive definite near the origin of the coupling constant space. It is however clear that in the absence of a theorem, the positivity of $\chi$ away from the origin is not guaranteed. The remaining coefficients of $\chi$ and $W$ are yet to be determined but, as we shall show, they are not needed to determine $\tilde{a}$ at the fixed points to the order investigated here.

\subsection{\texorpdfstring{Power of the consistency relations and the $\tilde{a}$-function}{Power of the consistency relations and the a-function}}

The system of first order differential equations in \eqref{eq:consistencycondition} allows  to derive the following conditions relating the different coefficients of the \betafunctions as well as $\chi$ and $W$,
\begin{eqnarray}
	\chi_{gg} b_y = -\chi_{yy} c_2 \ ,
	\quad\quad
	\chi_{yy} c_6 & = & \chi_{\lambda\lambda} d_3 \ ,
	\quad\quad
	4 \chi_{yy} c_7 = \chi_{\lambda\lambda} d_2 \ , \nonumber\\ 
	2 \chi_{gg} b_4 + \chi_{yy} \left( B_4 c_1 + B_3 c_2 + c_3 \right)
		& = & 2 \left( B_0 - C_2 + C_4 \right) c_1 \ ,\label{eq:relations} \\
	\chi_{gg} \left( B_2 b_0 + A b_y + b_3 \right)
		+ 2 \chi_{yy} \left(B_4 c_2 + c_4 \right)
		& = & \left( B_0 - C_2 + C_4 \right) c_2
			+ 2 \left( B_0 + C_2 - C_4 \right) b_0 \ . \nonumber
\end{eqnarray}
The equations in the first line above can be used to either test or determine some of the higher order coefficients of the \betafunctions since we know the metric factors. The remaining equations can be used in a similar fashion. However, given that the $B_i$ coefficients have not been explicitly computed  we use the knowledge of the \betafunctions to deduce, for example, $B_3$ and $B_4$ assuming that $c_2$ does not vanish.

For  $\tilde{a}$  to cubic order in the couplings, and using the consistency relations above, we have
\begin{equation}
	\tilde{a} = \afree + \tilde{a}^{(1)} + \tilde{a}^{(2)} + \tilde{a}^{(3)}  + \ldots \ ,
\end{equation}
where $\afree$ is the free--field theory value (\ref{eq:afree}), and the one, two and three--loops coefficients are \begin{eqnarray}
	\tilde{a}^{(1)} & = & -2 \chi_{gg} b_0 \alpha_g \ ,
	\label{eq:atilde:1} \\
	\tilde{a}^{(2)} & = & - \chi_{gg} \left( b_1 + A b_0 \right) \alpha_g^2
		- 2 \chi_{gg} b_y \alpha_g \alpha_y + \chi_{yy} c_1 \alpha_y^2 \ ,
	\label{eq:atilde:2} \\
	\tilde{a}^{(3)} & = & -\chi_{gg} \left[ \frac{2}{3} \left( b_2 + A b_1 \right) \alpha_g^3
		+ \left( b_3 + A b_y \right) \alpha_g^2 \alpha_y + 2 b_4 \alpha_g \alpha_y^2 
		+ \frac{1}{3} \frac{c_1}{c_2} \left( 4 b_4 - \frac{c_1}{c_2}
			\left( b_3 + A b_y \right) \right) \alpha_y^3 \right] \nonumber \\
		&& + \chi_{yy} \left[ \frac{2}{3} \left( c_5 - \frac{c_1}{c_2} c_3
			+ \left( \frac{c_1}{c_2} \right)^2 c_4 \right) \alpha_y^3
		+ c_6 \alpha_y^2 \alpha_\lambda + 2 c_7 \alpha_y \alpha_\lambda^2 \right]
		+ \frac{1}{3} \chi_{\lambda\lambda} a_1 \alpha_\lambda^3
		\label{eq:atilde:3} \\
		&& + \frac{\beta_{\alpha_g}}{\alpha_g^2} f\left( \alpha_i^3 \right)
		+ \frac{\beta_{\alpha_y}^2}{\alpha_y} \frac{B_0 - C_2 + C_4}{4 c_2} \ , \nonumber
\end{eqnarray}
where we defined
\begin{equation}
	f\left( \alpha_i^3 \right) = \chi_{gg} \left( \frac{B_1}{3} \alpha_g^3
		+ \frac{B_2}{2} \alpha_g^2 \alpha_y
		- \frac{B_2}{6} \left( \frac{c_1}{c_2} \right)^2 \alpha_y^3 \right)
		+ \frac{B_0 + C_2 - C_4}{3} \left( \frac{c_1}{c_2} \right)^2 \alpha_y^3 \ .
\end{equation}
Remarkably, the unknown coefficients $B_i$ and $C_i$ appear only in the last two terms of $\tilde{a}^{(3)}$, where they are multiplied by \betafunctions and hence vanish at fixed points\footnote{We have taken the liberty of adding higher order terms in order to rewrite the coefficients as \betafunctions. These terms are irrelevant when computing $\tilde{a}$ between perturbative fixed points.}. This was also observed to occur in supersymmetric theories~\cite{Freedman:1998rd}.

It is instructive to calculate $a$ to the second order using \eqref{eq:atilde} and recalling that the leading coefficients entering the one--form $W$ are $D_0 = \chi_{gg}$, $D_1 = \frac{1}{2} A \chi_{gg}$ and $D_2 = \frac{1}{2} \chi_{yy}$~\cite{Jack:1990eb}.  We find the simple expression 
\begin{equation}
	a = \afree + \chi_{gg} \left( b_1 \alpha_g^2
		+ b_y \alpha_g \alpha_y \right) + \mathcal{O}\left( \alpha_i^3 \right) \ ,
\end{equation}
where, remarkably, the term linear in $\alpha_g$, the one quadratic in $\alpha_y$ as well as the term linear in $A$ canceled out. Because the signs of $b_1$ and $b_y$ depend on the gauge theory, $a$ is not generally a monotonically decreasing function along the perturbative RG flow.

\subsection{\texorpdfstring{$\tilde{a}$ at fixed points}{a at fixed points}}

We can now move to the study of the fixed points and determine the variation of $\tilde{a}$ between two of them. A convenient way to search for the zeros of the system of \betafunctions \eqref{eq:betagauge}-\eqref{eq:betaquartic} is to first solve analytically for $\beta_{\alpha_\lambda} = 0$  which permits to relate $\alpha_\lambda$ to $\alpha_y$, then set to zero $\beta_{\alpha_y}$ further relating $\alpha_y$ to $\alpha_g$, so that finally we can search for the zeros of the following effective \betafunction in $\alpha_g$ 
\begin{equation}
	\beta_{\alpha_g}^\eff = -2\alpha_g^2 \left[b_0 + b_1^\eff \alpha_g + b_2^\eff \alpha_g^2 \right],
	\label{effbeta}
\end{equation}
where 
\begin{eqnarray}
	b_1^\eff & = & b_1 - \frac{c_2}{c_1} b_y \ , \label{eq:beta1eff} \\
	b_2^\eff & = & b_2 - \frac{c_2}{c_1} b_3 +
		\left(\frac{c_1}{c_2}\right)^2 b_4 - \frac{b_y}{c_1}
		\left[ c_4 - \frac{c_2}{c_1} c_3 + \left(\frac{c_2}{c_1}\right)^2 c_5^\eff \right] \ ,
	\label{eq:beta2eff}
\end{eqnarray}
with
\begin{equation}
	c_5^\eff = c_5 - \frac{d_3}{d_1} c_7 - \frac{d_2}{2 d_1}
		\left( c_6 - \frac{d_2}{d_1} c_7 \right)
		\left( 1 \pm \sqrt{1 - \frac{4 d_1 d_3}{d_2^2}} \right) \ .
\end{equation}
At this order in perturbation theory, there can be at most two perturbative fixed points for each sign in $c_5^\eff$, if both $b_0$ and $b_1^\eff$ are tuned to be small. An example of this is provided in the following section. Using eq.~(\ref{eq:beta1eff}) and (\ref{eq:beta2eff}), the difference in the function $\tilde{a}$ --- or equivalently $a$ --- between the UV and IR fixed points can then be written as
\begin{eqnarray}
	\Delta\tilde{a}_\textrm{perturbative} \equiv
		(\tilde{a}^{UV}-\tilde{a}^{IR})_\textrm{perturbative}
	= -2 \chi_{gg} && \left[ b_0 \left( \alpha_g^{UV}-\alpha_g^{IR} \right)
		+ \frac{1}{2} \left( b_1^\eff + A b_0 \right)
		\left( (\alpha_g^{UV})^2-(\alpha_g^{IR})^2 \right)  \right. \nonumber \\
	&& + \left. \frac{1}{3} \left( b_2^\eff + A b_1^\eff
		+ B b_0 \right) \left( (\alpha_g^{UV})^3-(\alpha_g^{IR})^3 \right) \right] \ ,
		\label{effa}
\end{eqnarray}
where $\alpha_g^{UV}$ and $\alpha_g^{IR}$ denote the values of the gauge coupling at the UV and IR fixed point respectively, and we defined
\begin{equation}
	B \equiv B_1 - \frac{c_2}{c_1} \left( B_2 + \frac{B_0}{\chi_{gg}} \right) \ .
\end{equation}
This expression reduces to the case of a gauge theory without Yukawa interactions by replacing $b_1^\eff$ and $b_2^\eff$ with $b_1$ and $b_2$.

Inspecting the effective \betafunction, there is a perturbative fixed point for small $b_0$, which reads 
\begin{equation}
	\alpha_g^{BZ}  = - \frac{b_0}{b_1^\eff} + \mathcal{O}\left( b_0^2 \right)\ .
\end{equation}
For positive $b_0$ and negative $b_1^\eff$, this is the usual Banks--Zaks IR fixed point. The situation in which the BZ fixed point is of UV nature is equally possible. This occurs by reverting the signs of both $b_0$ and $b_1^{\eff}$. The trivial fixed point at the origin will be in the first case an UV fixed point and in the second an IR one. The finite change in $\tilde{a}$ between the UV and IR fixed points can be computed either way, and one obtains
\ea{
 {\Delta\tilde{a}}^{BZ}=\mp \chi_{gg}\frac{b_0^2}{b_1^{\eff}}  \ .
 \label{BZda}
}
Here the sign reflects the sign of $b_0$, and $\Delta \tilde{a}$ is then positive for any physical fixed point. However, $\Delta \tilde{a}$ can formally become negative when the value of the coupling $\alpha_g$ at the fixed point is on the unphysical negative axis.

To three loop order in the effective \betafunction, one can have the two following physical zeros 
\begin{equation}
\label{merger}
\alpha_g^{BZ}=-\frac{b_1^\eff}{2b_2^\eff}\bigg(1 -  \sqrt{1-\frac{4b_0 b_2^\eff}{(b_1^\eff)^2}}\bigg) \ , \qquad \alpha_g^\nonBZ=-\frac{b_1^\eff}{2b_2^\eff}\bigg(1+ \sqrt{1-\frac{4b_0 b_2^\eff}{(b_1^\eff)^2}}\bigg)  
\end{equation}
For small values of $b_0$, the solution with negative sign corresponds to the usual BZ fixed point, with the following corrections
\begin{equation}
	\alpha_g^{BZ} =  - \frac{b_0}{b_1^{\eff}} \left( 1 + \frac{b_0 b_2^\eff}{{(b_1^\eff})^2}
		+ \mathcal{O}\left( b_0^2 \right) \right) \ .
	\label{eq:BZcoupling}
 \end{equation}
This expression holds provided $b_0/(b_1^\eff)^2$ is small. We shall see below that there are cases where this is not true. Using \eqref{eq:BZcoupling}, we can compute the three loop corrections to the variation of $\tilde{a}$, 
\ea{
 \Delta\tilde{a}^{BZ}=\mp {\chi_{gg}} \frac{b_0^2}{b_1^{\eff}} \left( 1 -  {(A b_1^\eff - 2 b_2^\eff)} \frac{b_0}{3(b_1^\eff)^2}\right)  \ .
}
We now turn our attention to the second zero $\alpha_g^\nonBZ$. The first observation is that for a generic value of $b_1^\eff$ this fixed point occurs at a non--perturbative value of the coupling. This is what happens in general for gauge theories with fermionic matter in a given irreducible representation of the gauge group \cite{Pica:2010xq}. However, for gauge theories with Yukawa interactions and/or multiple matter representations, the possibility that both $b_0$ and $b_1^\eff$ are small exists. An explicit example is provided below. Furthermore, when $b_2^\eff = (b_1^\eff)^2/(4b_0)$ the two fixed points merge. This phenomenon can happen within the range of perturbation theory. At the merger, one has $\alpha_\textrm{merger} = - 2 b_0 / b_1^\eff$, which, when plugged into Eq.~\eqref{effa}, gives
\ea{
 {\Delta\tilde{a}}^{BZ}\big|_\textrm{merger}=\mp \chi_{gg} \frac{4}{3} \frac{b_0^2}{b_1^{\eff}}  \ .
}
The virtues of this expression will be studied in more detail elsewhere~\cite{Antipin:2013yyy}.

Having in our hands the explicit tools, we can explore the \atheorem for gauge theories with interesting fixed point structures.

\section{A concrete example}
\label{example}

We consider a $SU(N_c)$ gauge theory with $N_f$ fundamental Dirac fermions  $Q = (q, \widetilde{q}^*)$, $\ell$ adjoint Weyl fermions $\lambda$, and a gauge singlet complex scalar $H$ that transforms in the bifundamental representation of the $SU(N_f)_L \times SU(N_f)_R$ global symmetry of the theory. For the benefit of the reader, the field content and the quantum symmetries of the theory with $\ell =1$ are summarized in Table~\ref{FieldContent}. The Lagrangian of the theory is
\ea{
\mathcal{L}&= \tr \left[- \frac{1}{2} F^{\mu \nu}F_{\mu \nu} +i \bar{\lambda}  \slashed{D} \lambda+ \overline{Q} i \slashed{D} Q + \partial_\mu H ^\dagger \partial^\mu H + y_H \overline{Q} H Q  \right]  - u_1 (\tr [H ^\dagger H])^2 -u_2\tr(H ^\dagger H )^2 \ . 
 \label{eq:Llsm}
}
$\tr$ is the trace over both color and flavor indices. ${D_\mu}$ is the usual covariant derivative.
\begin{table}[b]
\vspace{3mm}
\caption{Field content of the example. The first three fields are Weyl spinors in the ($\frac{1}{2},0$) representation of the Lorentz group. $H$ is a complex scalar and $G_\mu$ is the gauge field. $U(1)_{AF}$ is the extra Anomaly Free symmetry
arising due to the presence of $\lambda$.}%
\vspace{-5mm}
\[ \begin{array}{c|c|c c c c} \hline \hline
{\rm Fields} &\left[ SU(N_c) \right] & SU(N_f)_L &SU(N_f)_R & U(1)_V& U(1)_{AF} \\ \hline 
\lambda & {\rm Adj} & 1 & 1 & 0 & 1 \\
 q &\yf &\overline{\yf }&1&~~\frac{N_f-N_c}{N_c} & - \frac{N_c}{N_f}  \\
\widetilde{q}& \overline{\yf}&1 &  {\yf}& -\frac{N_f-N_c}{N_c}& - \frac{N_c}{N_f}     \\
 \hline
  H & 1 & \yf & \overline{\yf} & 0 & \frac{2N_c}{N_f}\\
  G_\mu & \text{Adj} & 1 & 1 & 0 & 0 \\
   \hline \hline \end{array}%
\]%
\label{FieldContent}%
\vspace{-5mm}
\end{table}
Throughout this section we will work with the rescaled couplings which enable a finite Veneziano limit of the theory with $\ell$ fixed. That is, we let $N_c, N_f \to \infty$ while keeping $x \equiv N_f/N_c$ fixed. The  appropriately rescaled couplings are
\ea{
a_g = \frac{g^2N_c}{(4 \pi)^2} \ ,~\ a_H = \frac{y_H^2N_c}{(4 \pi)^2}\ ,~ z_1 = \frac{u_1N_f^2}{(4 \pi)^2}\  , ~z_2 = \frac{u_2N_f}{(4 \pi)^2} \ .
}
This model was introduced in \cite{Antipin:2011aa} to investigate near--conformal dynamics, at the one loop level, and its impact on the spectrum of the theory with special attention to the dilaton properties. The model was further investigated at the two loop level in \cite{Antipin:2012kc}. To compute $\tilde{a}$, following the previous section, we need to determine the three loop contribution to the gauge \betafunction.

Using \cite{Machacek:1983tz, Machacek:1983fi, Machacek:1984zw, Pickering:2001aq, Mihaila:2012pz,Harlander:2009mn} we find in the Veneziano limit 
\ea{
\label{betaag}\beta_{a_g} ={}& -\frac{2}{3} a_g^2 \left[11-2\ell-2 x+\left(34-16\ell-13 x\right)a_g +3x^2 a_H \right. \nonumber \\
&+\left. \frac{81x^2}{4}a_g a_H-\frac{3x^2(7+6x)}{4}a_H^2+\frac{2857+112x^2-x(1709-257 \ell)-1976 \ell +145\ell^2}{18}a_g^2  \right] \ , \\[2mm]
\label{beta1H}\beta_{a_H} ={}& a_H  \left[ 2(x+1) a_H - 6 a_g +(8 x+5) a_g a_H+\frac{20 (x+\ell)-203}{6} a_g^2-8 x z_2 a_H-\frac{x(x+12)}{2} a_H^2+4 z_2^2\right] \ , \\
\label{beta2Y} \beta_{z_2} ={}& 2 \left(2 z_2 a_H+4 z_2^2-x a_H^2\right) \ .
}
Here one can see that the double trace coupling $z_1$ does not participate in the running of the remaining couplings. In addition, using \eqref{eq:chis} and \eqref{eq:A} the metric coefficients for this theory can be found:
\begin{equation}
	\chi_{gg} = \frac{N_c^2 }{2^7 \pi^2} \ ,
	\quad
	\chi_{yy} = \frac{N_f^2}{3 \cdot 2^7 \pi^2} \ ,
	\quad
	\chi_{\lambda\lambda} = \frac{N_f^2}{3 \cdot 2^6 \pi^2} \ ,
	\quad
	A = 17-\frac{10}{3}(x+\ell)  \ .
	\label{metricAMS}
\end{equation}
One can check that the expressions above satisfy the consistency relations given in the first line of~\eqref{eq:relations}, and therefore it constitutes an independent check of the correctness of the \betafunctions.  We now turn to the FP analysis of the model which will reveal an interesting perturbative structure.   

\subsection{Leading order analysis: Banks--Zaks fixed point}

In order to see a physical BZ fixed point, the one--loop coefficient of the gauge \betafunction has to be small and the signs of $b_0$ and $b_1^{\eff}$ have to be opposite. Therefore, our first task is to find a region in the parameter space of the model where the physical BZ fixed point exists. We use Eq.~\eqref{eq:beta1eff}
\begin{equation}
\label{tuning}
b_0= \frac{1}{3}\big(11-2(\ell+x)\big) \ ,  \qquad  b_1^{\eff}=\frac{1}{3}\big(34-16\ell-13x+\frac{9x^2}{(x+1)}\big) \ .
\end{equation}
From the asymptotic freedom (AF) boundary condition $b_0=0$ we obtain that $x=(11-2\ell)/2$. Substituting this value of $x$ into $b_1^{\eff}$, we have
\begin{equation}
\label{tuning1}
b_{1AF}^{\eff}=-\frac{25}{2}-\ell-\frac{3(11-2\ell)^2}{4\ell-26}\ ,
\end{equation}
where the last term comes from the Yukawa interactions. We immediately notice that for $\ell^*\approx 0.37$ the $b_{1AF}^{\eff}$ vanishes. Below, we will consider the cases $\ell =1 $ for which this coefficient is negative and $\ell=0$ for which it is positive. In the first case we have a standard IR BZ fixed point, and in the second we obtain a new UV BZ fixed point. It is worth noticing that, in the absence of Yukawa interactions, $b_{1AF}^{\eff}$ in \eqref{tuning1} is always negative and therefore the physical BZ FP can only be the standard IR fixed point. 

\subsubsection{\texorpdfstring{$\ell=1$ case}{l=1 case}}
In this case there exists a perturbative IR fixed point regardless of whether we consider the presence of Yukawa interactions.  We show  in Fig.~\ref{fig:a:LO:1} the leading order result for the change in $\tilde{a}$ between the Gaussian (trivial) FP and the BZ IR one at leading order, both in the presence (blue line) and absence (red line) of Yukawa interactions. Both curves cross zero at  $x^*=9/2$, when asymptotic freedom is lost.  For $x>9/2$, where $b_0<0$, there is an unphysical BZ UV fixed point yielding a negative $\Delta\tilde{a}$. The Yukawa interactions in \eqref{tuning1} imply  $|b_1^{\eff}|<|b_1|$, which leads to a larger $\Delta \tilde{a}$ in the case of the gauge theory with scalars. 
\begin{figure}[bt]
\subfloat[$\Delta \tilde{a}$ between the Gaussian and BZ fixed points normalised to $\chi_{gg}$ at leading order for the $\ell=1$ case. The red (dashed blue) line corresponds to the model without (with) Yukawa interactions. In both cases the physical BZ FP is an IR one.]{\includegraphics[width=.47\textwidth]{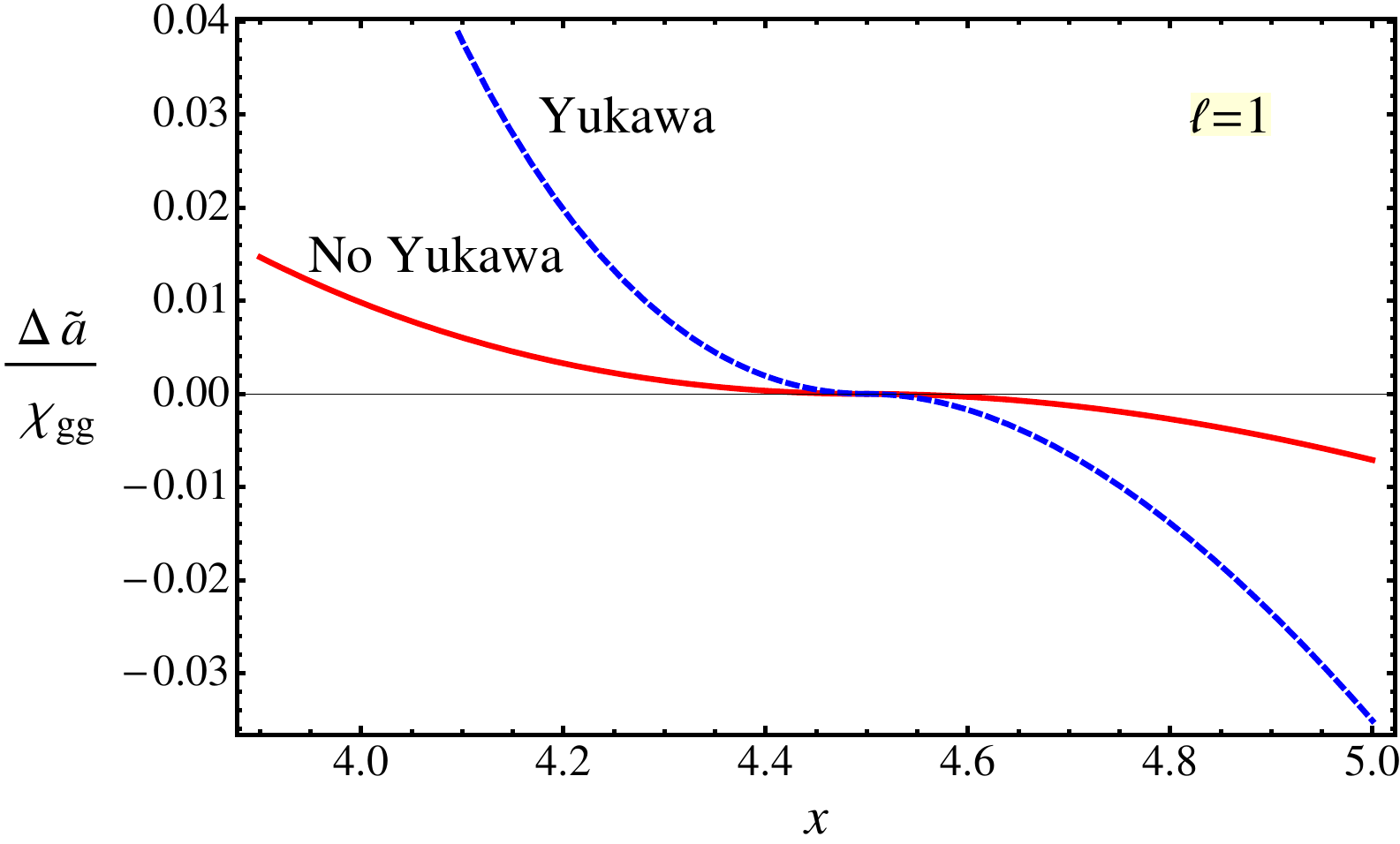}\label{fig:a:LO:1}}
  \hfill
\subfloat[$\Delta \tilde{a}$ between the Gaussian and BZ fixed points normalised to $\chi_{gg}$ at leading order for the $\ell=0$ case. In the absence (presence) of Yukawa interactions, the physical BZ fixed pont is an IR (UV) one. The color code is the same as on the left panel.]{\includegraphics[width=.47\textwidth]{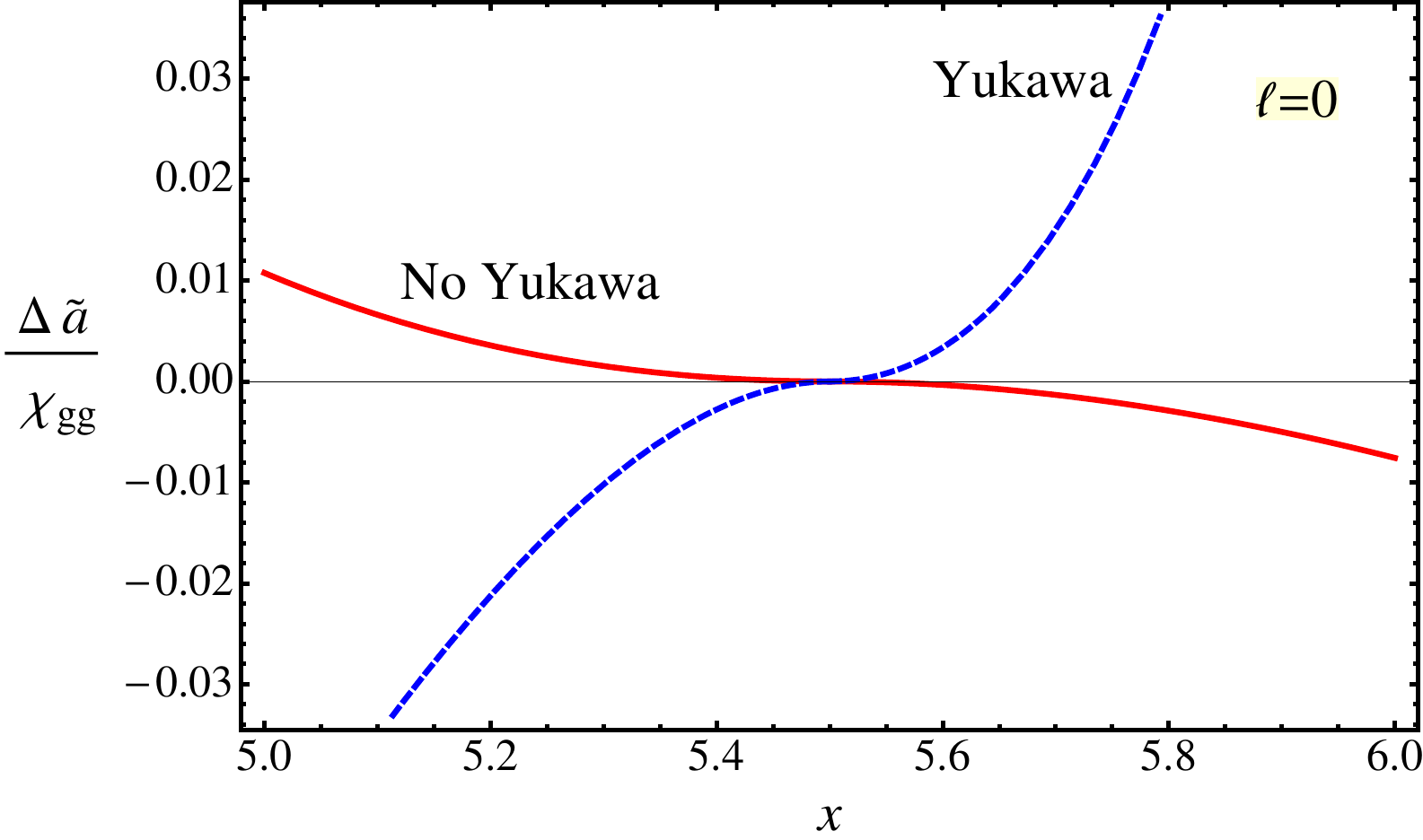}\label{fig:a:LO:0}}
\addtocounter{figure}{1}
\end{figure} 

\subsubsection{\texorpdfstring{$\ell=0$ case}{l=0 case}}
\label{ell=0}

We now turn to the $\ell=0$ case where, rather than having a BZ IR fixed point, the theory develops a UVFP when asymptotic freedom is lost, i.e. for  $b_0<0$. Of course, this is possible only because of the presence of the Yukawa interactions. In Fig.~\ref{fig:a:LO:0},  the leading order result for the change in $\tilde{a}$ between the Gaussian and BZ fixed points without Yukawa interactions is shown in red, and the one with Yukawa interactions in blue. Both curves cross zero for $x^*=11/2$ when asymptotic freedom is lost.

\subsection{Next--to--leading order analysis: Fixed point merger}

At the next perturbative order we deal with the full system of Eqs.~\eqref{betaag}--\eqref{beta2Y} and from now on, we concentrate only on the physical fixed points.

\begin{figure}[bt]
\subfloat[The next--to--leading order physical FP structure for the $\ell=1$ case with Yukawa and quartic interactions.] 
{\includegraphics[width=.48\textwidth]{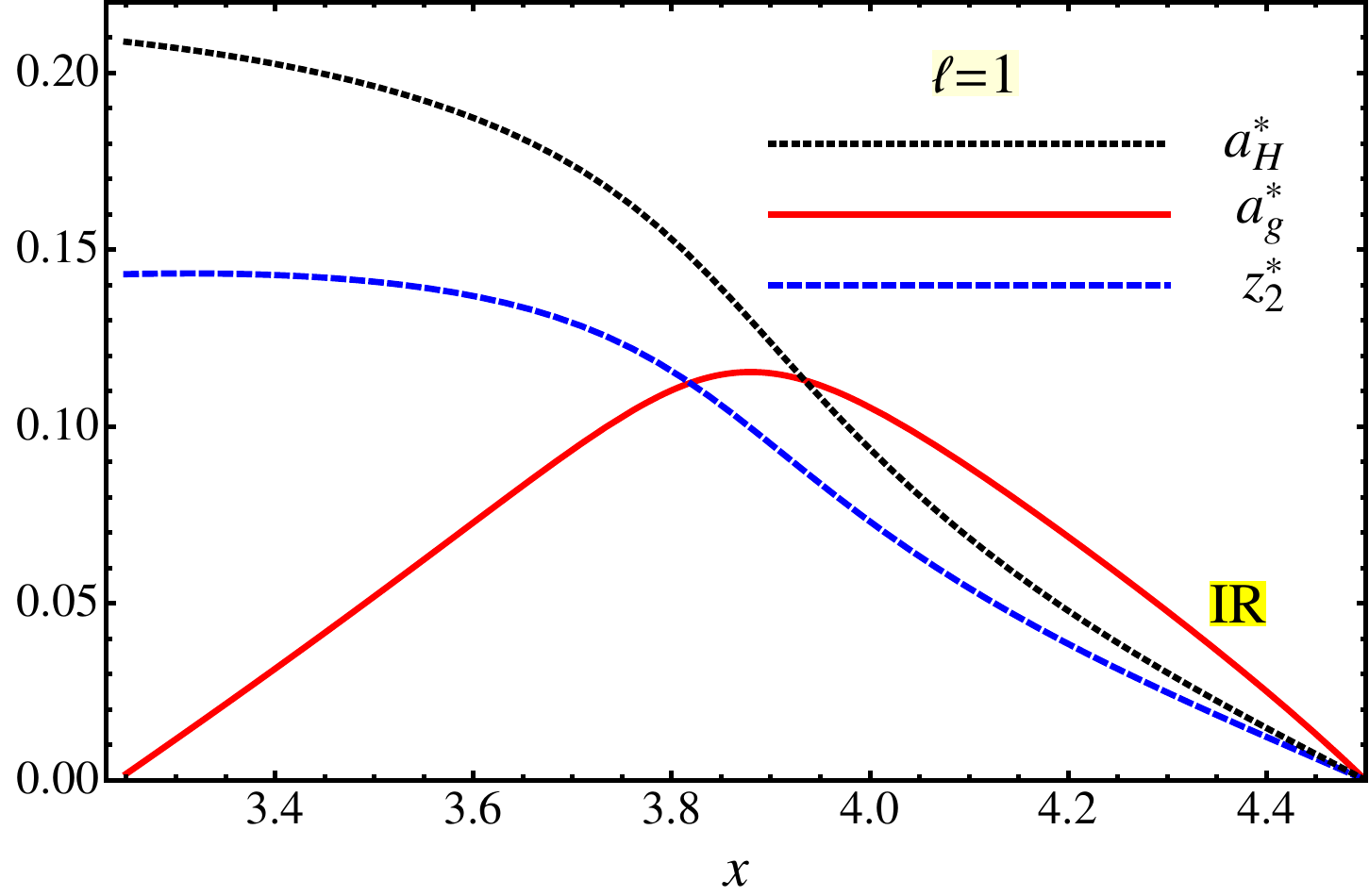}\label{fig:FP:1}}
  \hfill
\subfloat[$\Delta \tilde{a}$ normalised to $\chi_{gg}$ for the $\ell=1$ case. The red and dashed blue lines are leading order results from Fig.~\ref{fig:a:LO:1} while the dotted black and green lines are the next--to--leading order corrections.]{\includegraphics[width=.51\textwidth]{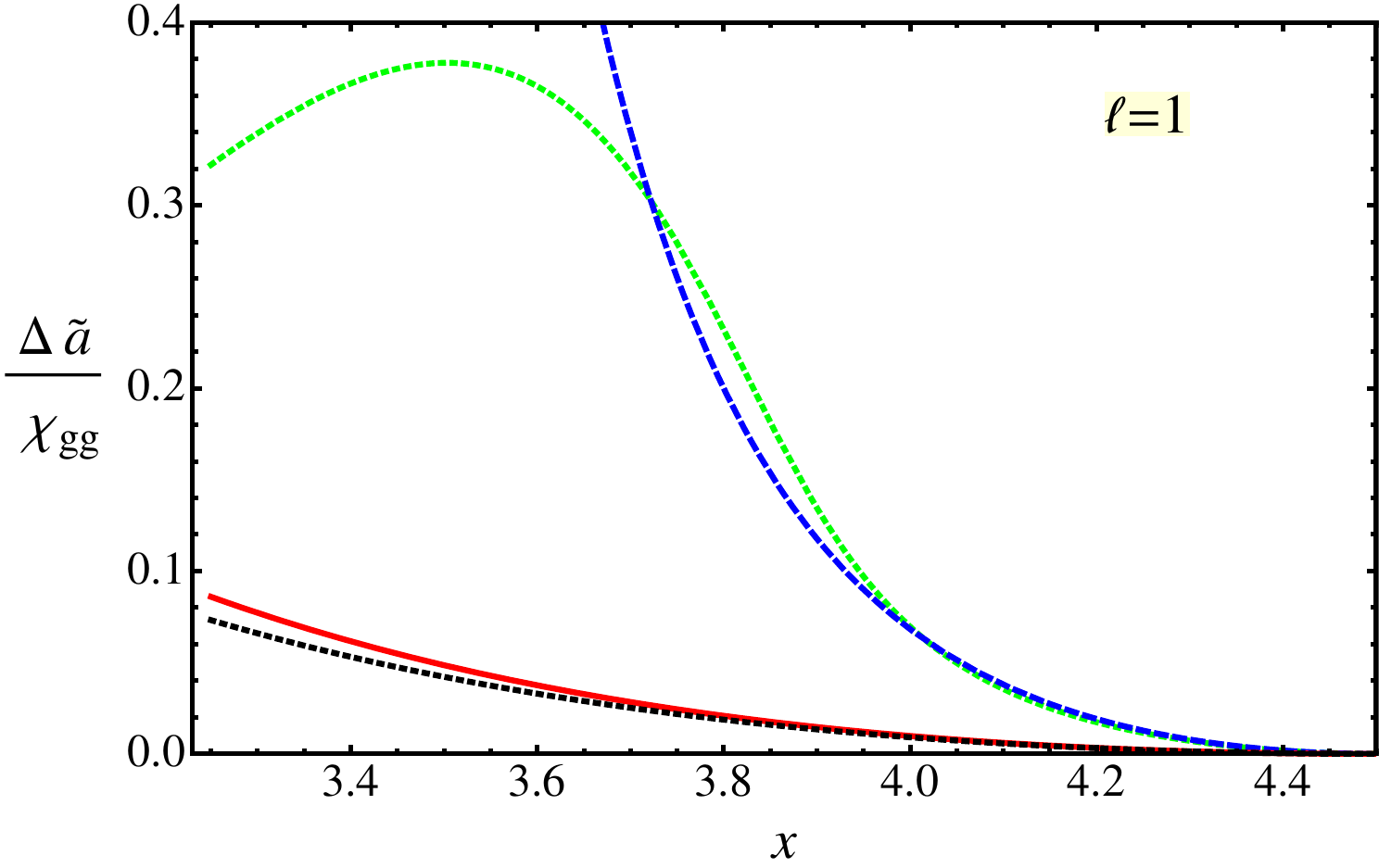}\label{fig:a:NLO:1}}
\addtocounter{figure}{1}
\end{figure} 

\subsubsection{\texorpdfstring{$\ell=1$ case}{l=1 case}}

We start again with the $\ell=1$ theory and in Fig.~\ref{fig:FP:1} we display the FPs structure for the model with Yukawa and quartic interactions. We notice that at $x\approx 3.25$ the FP value of the gauge coupling vanishes. However, this happens in the region beyond applicability of perturbation theory since the two remaining coupling constants are large. In Fig.~\ref{fig:a:NLO:1} we plot the change in the $\tilde{a}$--function for the next--to--leading order BZ IR fixed point and compare it with the corresponding leading order results from Fig.~\ref{fig:a:LO:1}. As a general feature, we notice that the next--to--leading order corrections reduce the value of $\Delta\tilde{a}$ in the perturbative regime. It is clear from the plots that for the theory with Yukawa interactions perturbation theory breaks down, when moving away from the critical value $x^* = 4.5$, earlier with respect to the theory without Yukawa interactions.

\subsubsection{\texorpdfstring{$\ell=0$ case}{l=0 case}}
 
\begin{figure}[bt]
\subfloat[The next--to--leading order physical FP structure for the $\ell=0$ case \emph{with} Yukawa and quartic interactions. The vertical dashed--dotted green line represents the point where asymptotic freedom is lost. ]{\includegraphics[width=.46\textwidth]{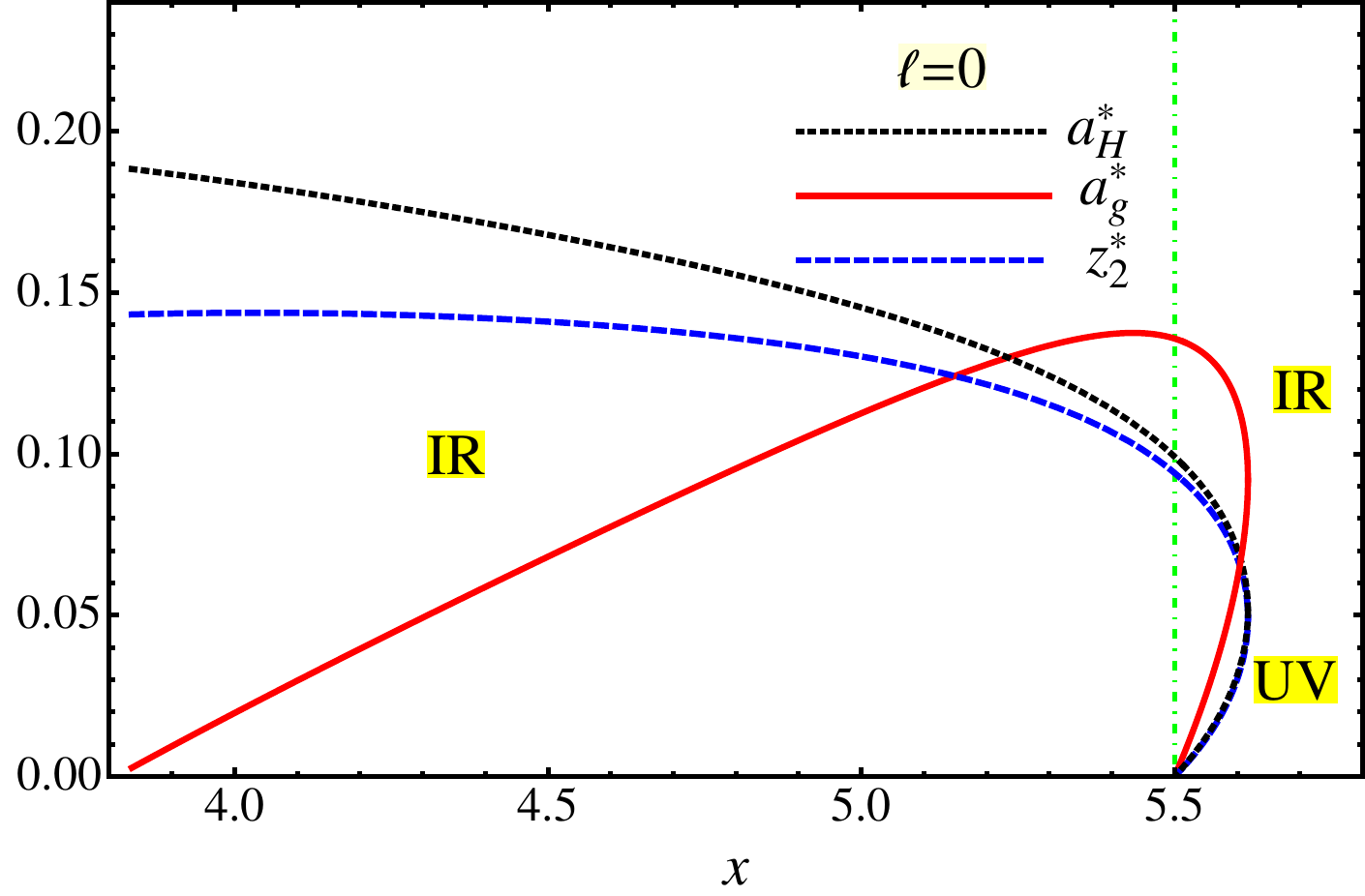}\label{fig:FP:0}}
  \hfill
\subfloat[$\Delta \tilde{a}$ normalised to $\chi_{gg}$ for  the $\ell=0$ case. The red and dashed blue lines are leading order results from Fig.~\ref{fig:a:LO:0} while the dotted black and green lines are next--to--leading order corrections.]{\includegraphics[width=.51\textwidth]{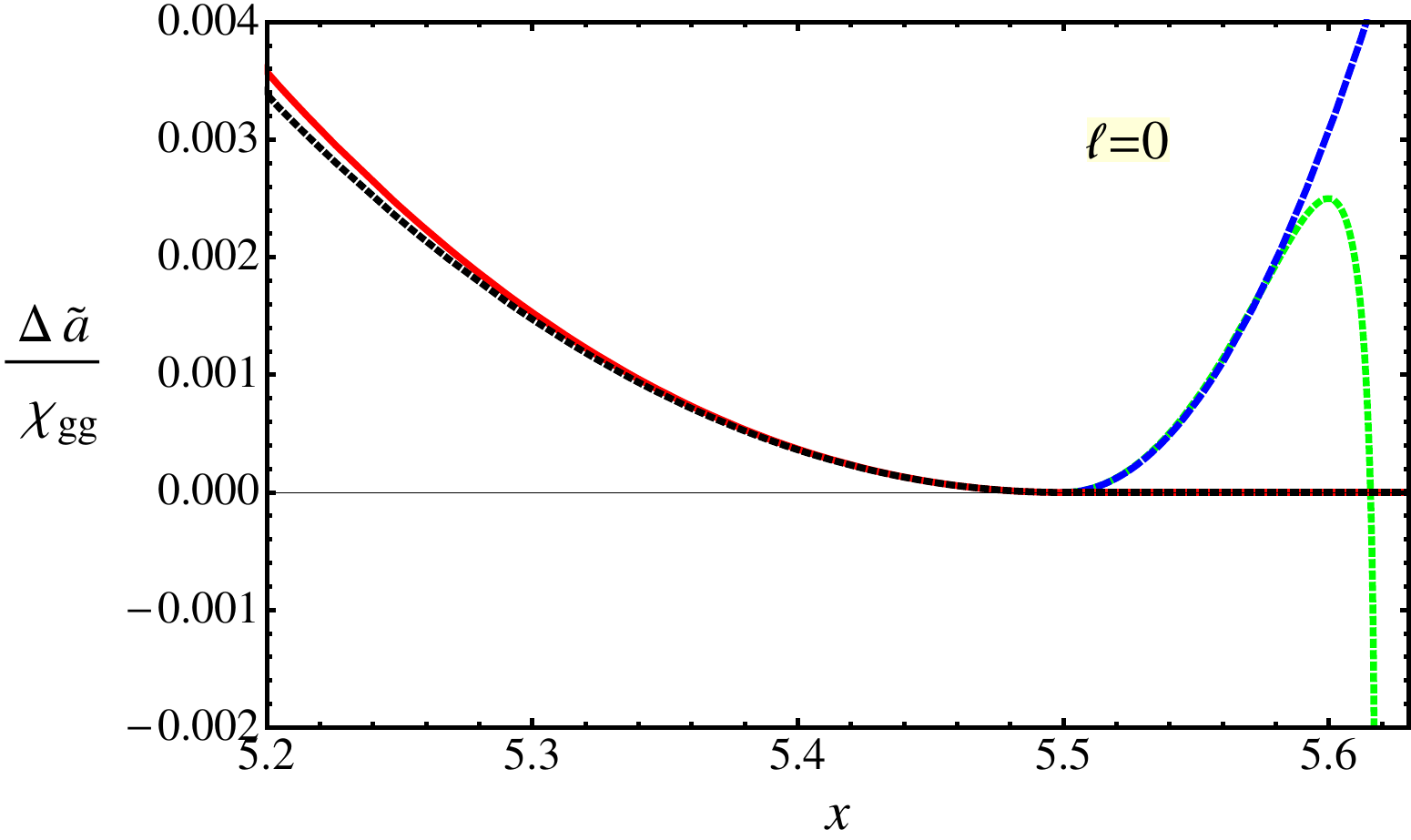}\label{fig:a:NLO:0}}
\addtocounter{figure}{1}
\end{figure} %
We now turn to the $\ell=0$ theory where, as discussed above, there is no BZ IR fixed point. However, in the asymptotically free regime there is a new physical IR fixed point emerging at the next--to--leading order. This non--BZ IR fixed point is present also when asymptotic freedom is lost, i.e. for $x>5.5$.  In this region there is also a BZ UV fixed point that was discussed in subsection~\ref{ell=0}. The complete FP structure is shown in Fig.~\ref{fig:FP:0}. The UV and IR fixed points merge around $x\approx 5.6$. In Fig.~\ref{fig:a:NLO:0} we plot the change in the $\tilde{a}$--function at next--to--leading order together with the corresponding leading order results from Fig.~\ref{fig:a:LO:0}. We notice that $\Delta\tilde{a}$ becomes negative just before the merger which is incompatible with the \atheorem. We interpret this effect as the breakdown of the perturbative expansion since the FP values of the couplings at the merger are quite large, as can be seen from Fig.~\ref{fig:FP:0}.

So far, all our calculations of $\Delta\tilde{a}$ were for the flow connecting the trivial FP at the origin of the coupling constant space with the BZ one. However, it is relevant also to determine $\Delta\tilde{a}$ for the branch connecting the two non--trivial FPs. In the theory with $\ell=0$ and $x>5.5$ this is the RG flow between the BZ UV fixed point and the non--BZ IR one. We display the change in the $\tilde{a}$--function in Fig.~\ref{fig:nonBZbranch}. Of course, at the merger $\Delta \tilde{a}$ vanishes.

\subsubsection{\texorpdfstring{$\ell=0.35$ case: The perturbative merger}{l=0.35 case: The perturbative merger}}

Using both $\ell$ and $x$ as continuous parameters it is formally possible to study the merging phenomenon within the perturbative regime. This happens around $\ell\approx 0.37$. Therefore we provide an example with $\ell=0.35$. The change in the $\tilde{a}$--function for the two RG flows are shown in Fig.~\ref{fig:bothbranches}. Since for this value of $\ell$ perturbation theory holds, we observe a positive and well behaved $\Delta \tilde{a}$ all the way to the merger.

\begin{figure}[bt]
\subfloat[$\Delta \tilde{a}$ normalised to $\chi_{gg}$ for the RG flow between the BZ UV fixed point and the non--BZ IR fixed point.]{\includegraphics[width=.47\textwidth]{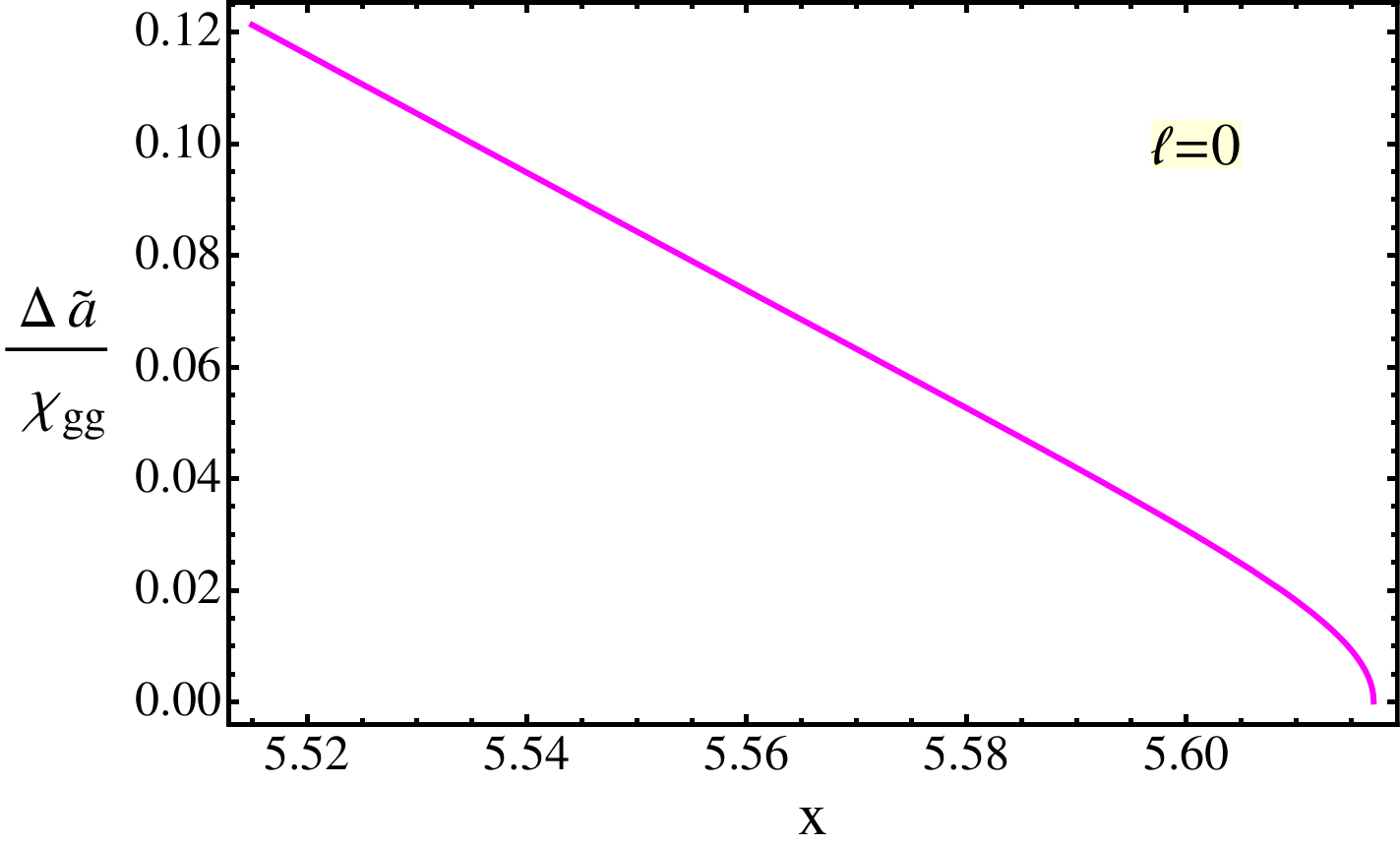}\label{fig:nonBZbranch}}
  \hfill
\subfloat[$\Delta \tilde{a}$ normalised to $\chi_{gg}$ for  the $\ell=0.35$ case. The magenta curve corresponds to the flow between the BZ UV and the non--BZ IR fixed points. The green curve is the result between the Gaussian IR fixed point and the UV BZ one.]{\includegraphics[width=.5\textwidth]{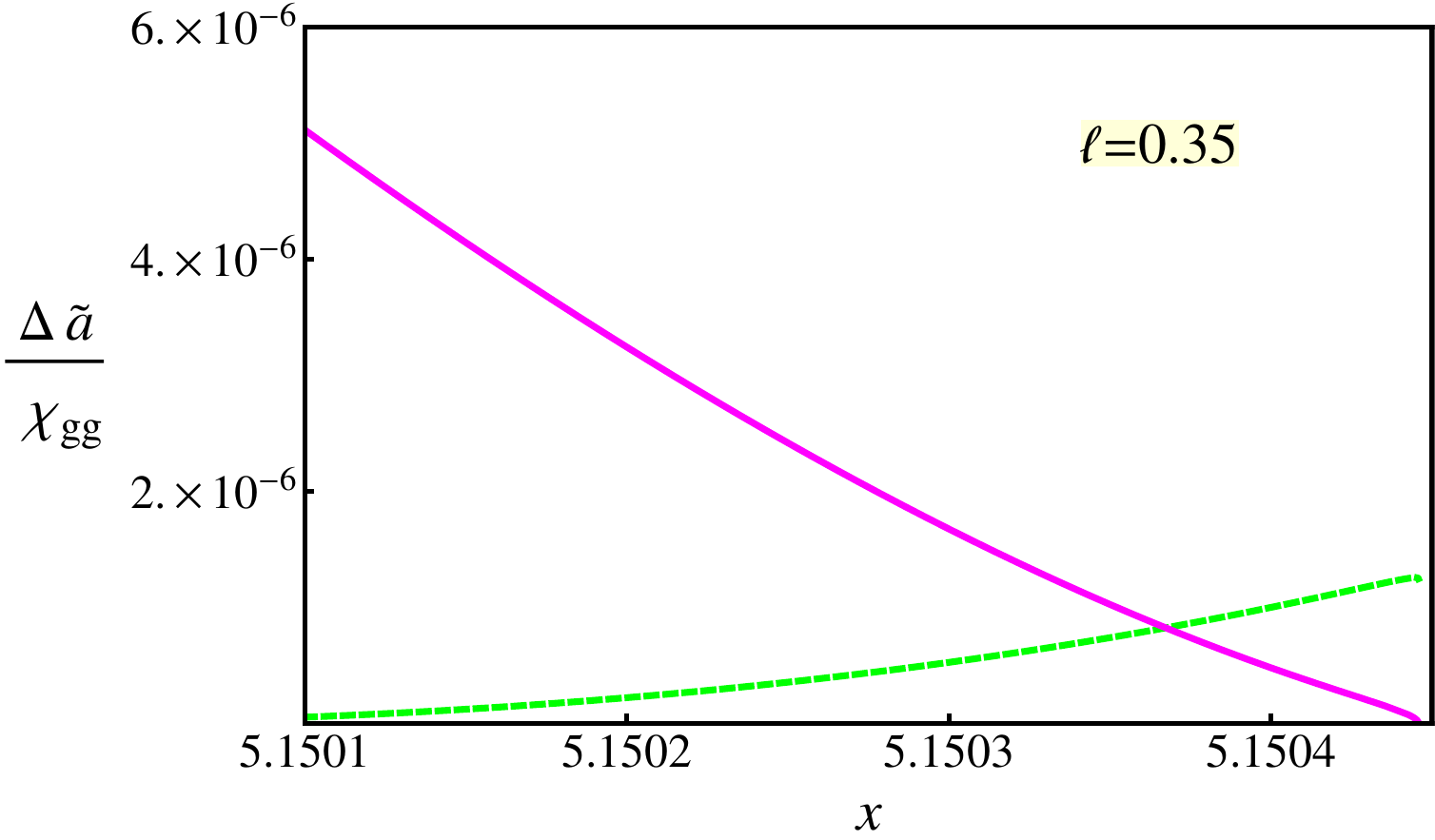}\label{fig:bothbranches}}
\addtocounter{figure}{1}
\end{figure} 
\section{Conclusions}
\label{conclusions}

We studied the \atheorem to the maximum known order in perturbation theory for nonsupersymmetric gauge theories with fermionic matter as well as gauge singlet scalar fields, where the latter two interact via Yukawa interactions. The computation involves three loops in the gauge \betafunction, two loops in the Yukawa interactions and one loop in the quartic interactions. To this order we have first determined the general expression for the change in the function $\tilde{a}$ between two fixed points, and then specialised it to the case of QCD with extra adjoint Weyl fermions and elementary mesons interacting via Yukawa terms. We employed the Veneziano limit and determined the three loop \betafunctions of the theory. This has allowed us to test the \atheorem beyond the lowest order in perturbation theory. We discovered that the model posses an interesting structure of the fixed points, depending on the number of adjoint fermionic matter fields, among which we highlight the presence of a Banks--Zaks UV fixed point as well as the occurrence of a merging phenomenon for which the \atheorem had not been tested before. 
 
\subsection*{Acknowledgements}
The CP$^3$-Origins centre is partially funded by the Danish National Research Foundation, grant number DNRF90. We thank John Gracey for clarifying comments regarding the calculation of the three loop gauge \betafunction. OA thanks the Galileo Galilei Institute for Theoretical Physics for the hospitality and the INFN for partial support during the completion of this work.

\appendix
\section{Diagrammatic form of the trace anomaly}
\label{appendix}

The structure of the metric (\ref{eq:metric}) and one--form (\ref{eq:oneform}) entering the trace anomaly can be determined by looking at vacuum polarisation diagrams containing in addition to the usual vertices of the quantum field theory the following counterterms
\begin{equation*}
	\begin{array}{cccccc}
		\includegraphics[width=2.5cm]{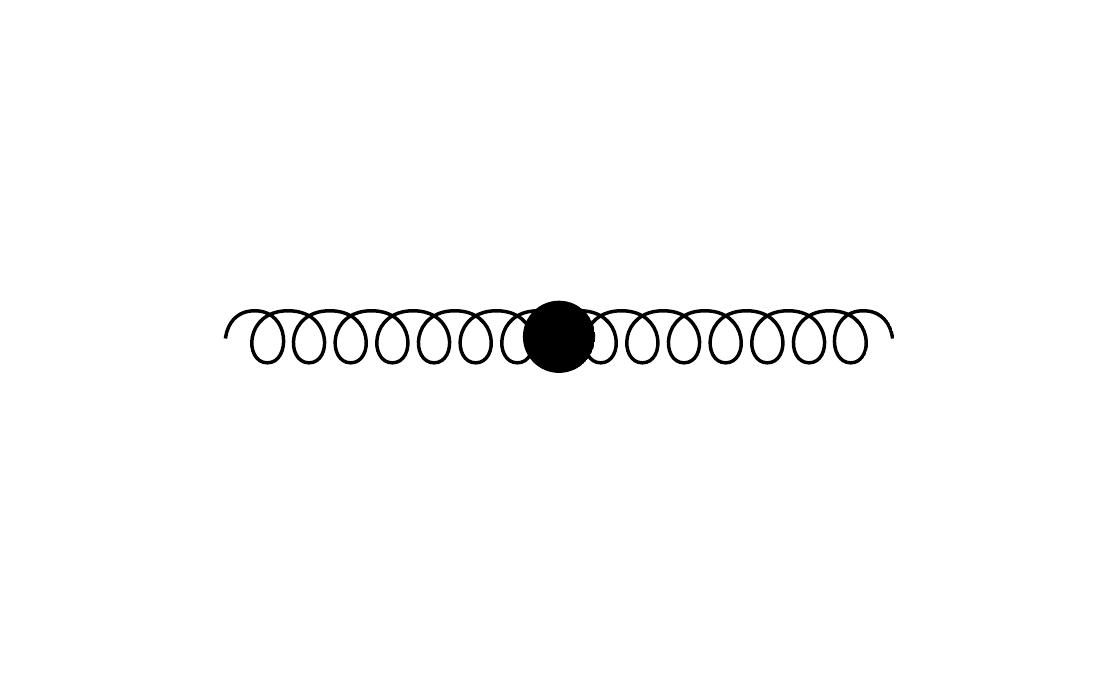} &
		\includegraphics[width=2.5cm]{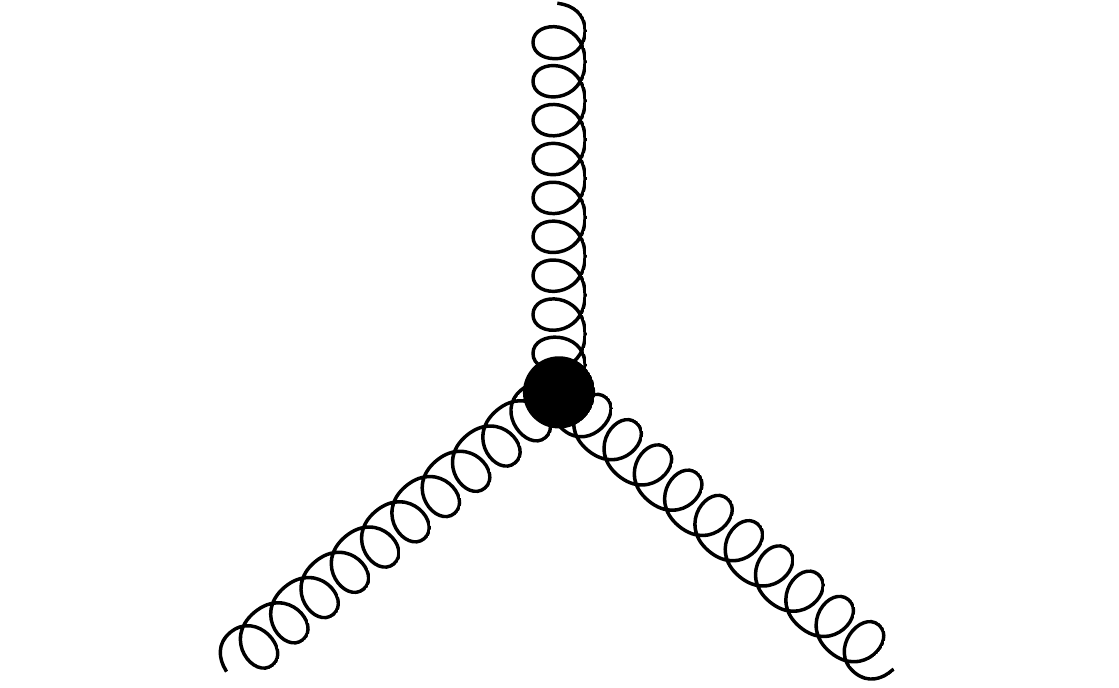} &
		\includegraphics[width=2.5cm]{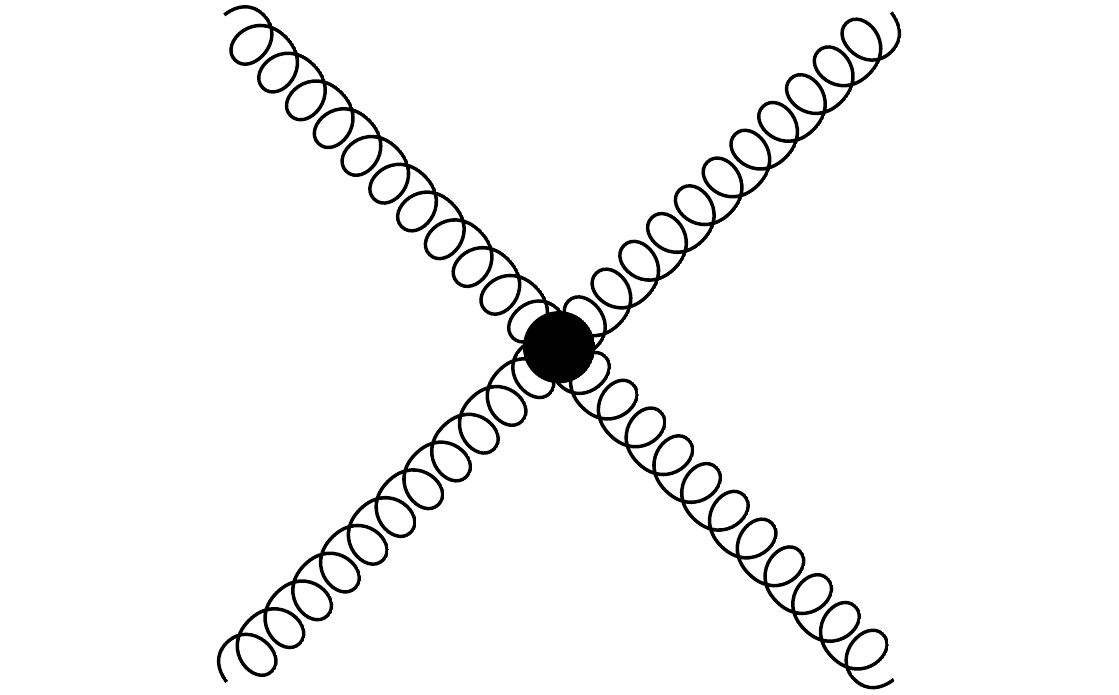} &
		\includegraphics[width=2.5cm]{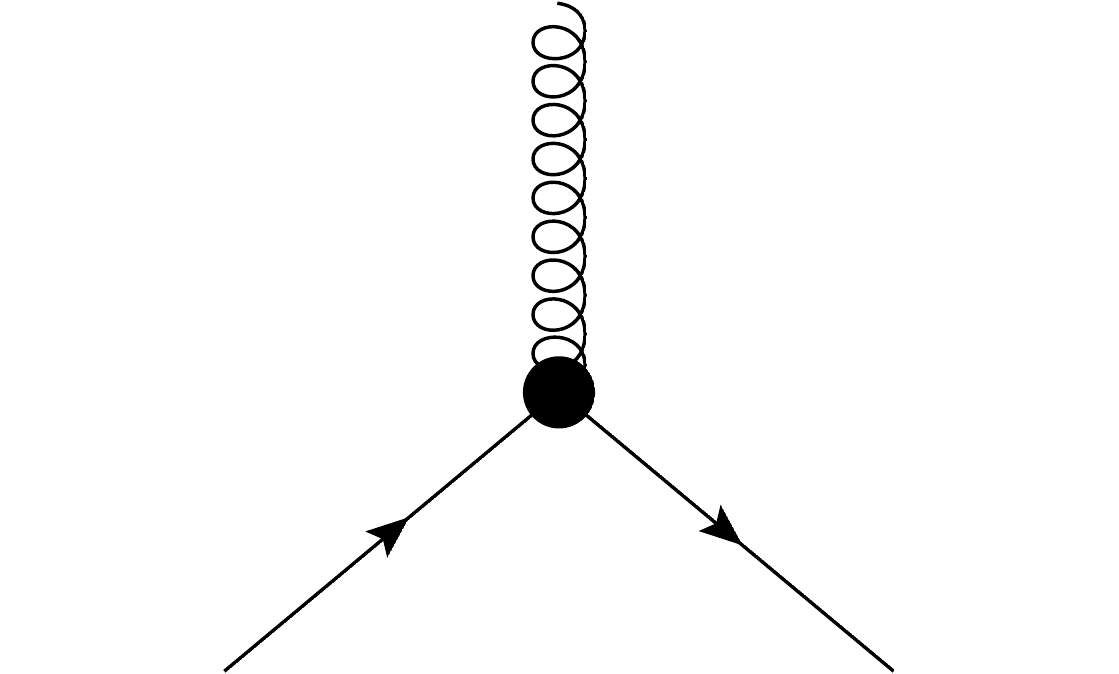} &
		\includegraphics[width=2.5cm]{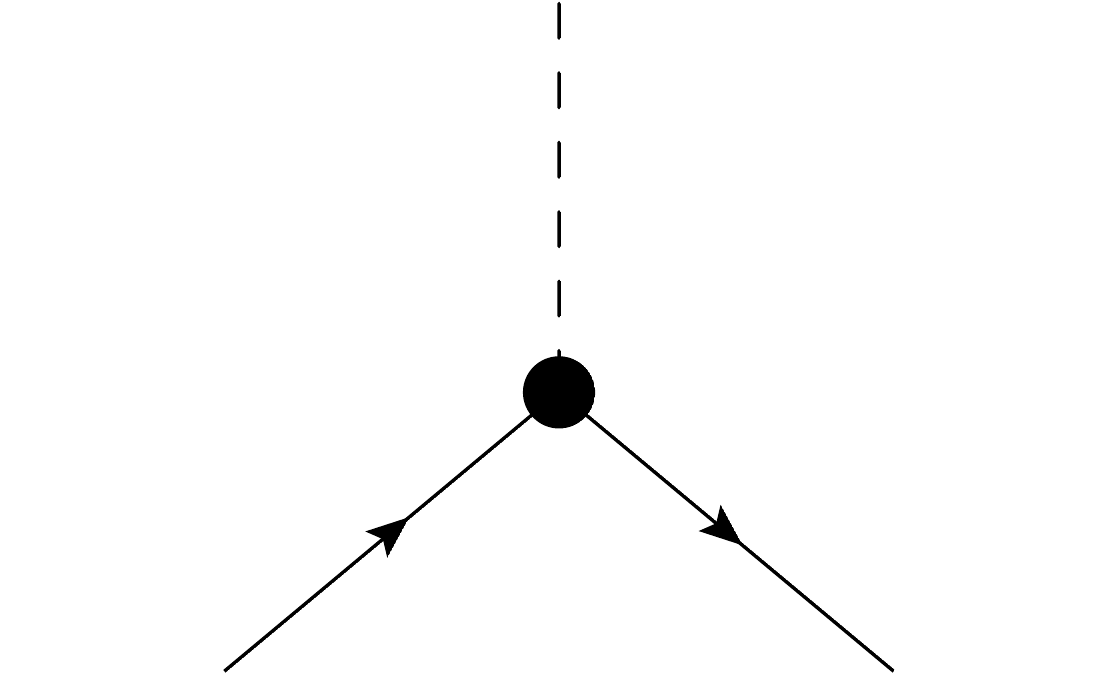} &
		\includegraphics[width=2.5cm]{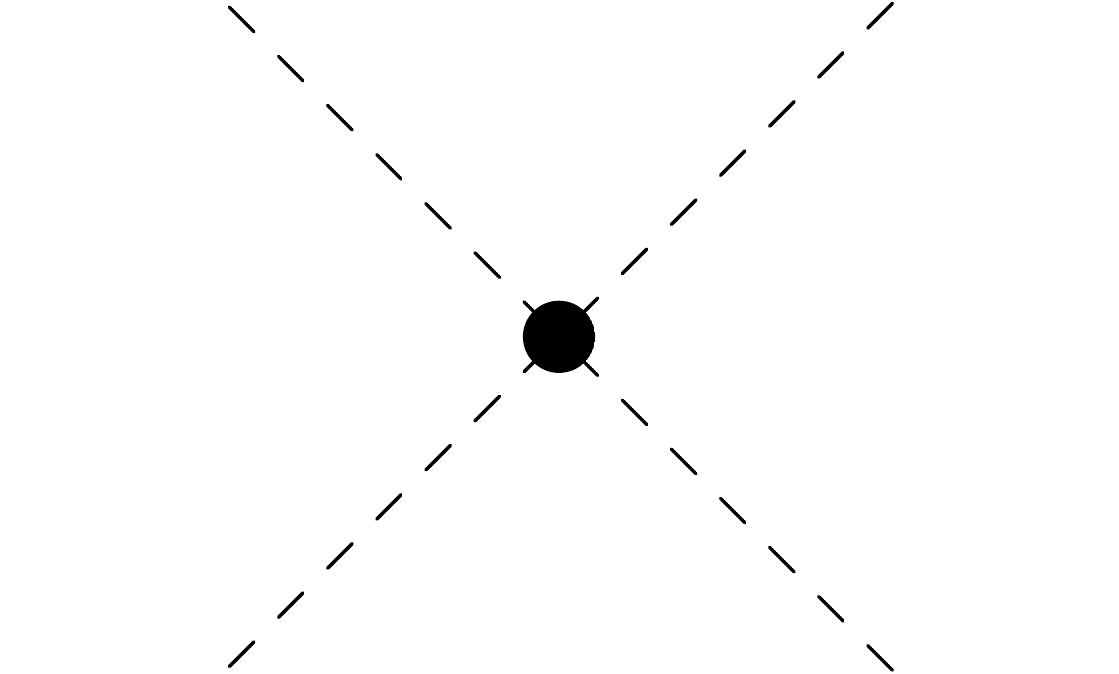} \\
		\frac{\beta_g}{g} \sim \frac{\beta_{\alpha_g}}{\alpha_g} &
			\beta_g \sim \frac{\beta_{\alpha_g}}{\alpha_g^{1/2}} &
			g \beta_g \sim \beta_{\alpha_g} &
			\beta_g \sim \frac{\beta_{\alpha_g}}{\alpha_g^{1/2}} &
			\beta_y \sim \frac{\beta_{\alpha_y}}{\alpha_y^{1/2}} &
			\beta_\lambda \sim \beta_{\alpha_\lambda}
	\end{array}
\end{equation*}
The terms entering the metric (\ref{eq:metric}) are the ones proportional to two powers of the \betafunctions while  the ones with one power of $\beta_i$ fix the one form $W$ (\ref{eq:oneform}). All the vacuum polarisation diagrams up to three--loop order as well as the form of their contribution are shown in Table~\ref{tab:vacpoldiagrams}. Note that more diagrams would be present if the scalar field were charged under the gauge group.

\clearpage
\begin{table}[h!t]
	\begin{tabular}{|c|c|c|}
		\hline
		Diagrams & Contributions to $\chi$ & Contributions to $W$ \\
		\hline\hline
		\raisebox{-0.6cm}{\includegraphics[width=2.5cm]{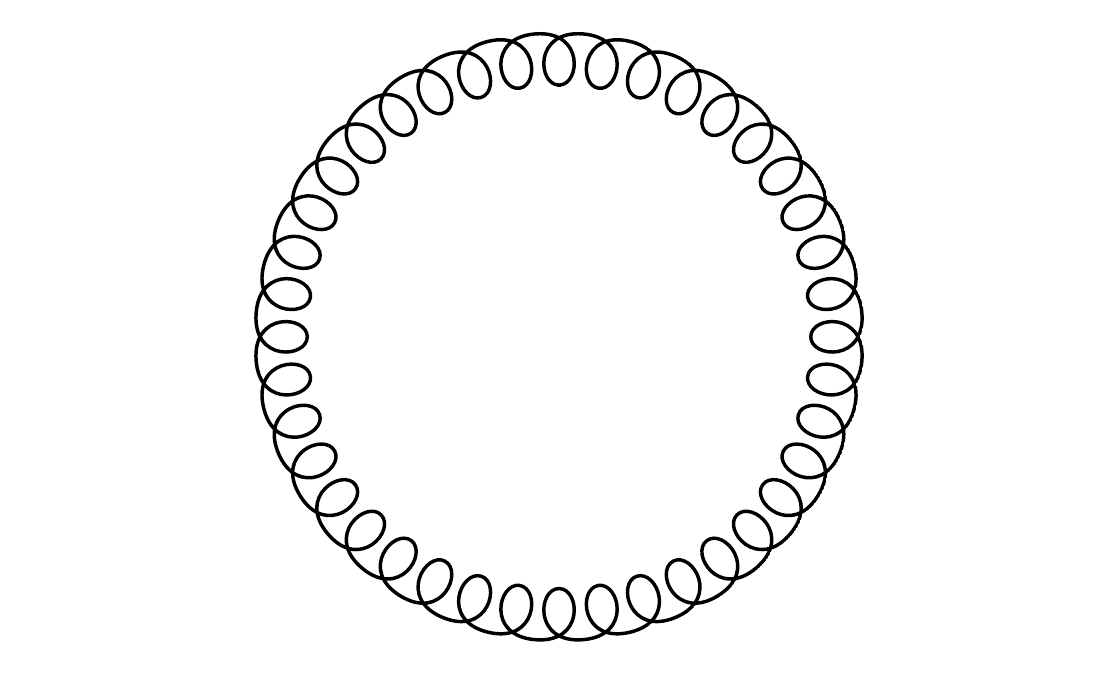}} &
			$\frac{\beta_{\alpha_g}^2}{\alpha_g^2}$ & $\frac{\beta_{\alpha_g}}{\alpha_g}$ \\
		\hline\hline
		\raisebox{-0.6cm}{\includegraphics[width=2.5cm]{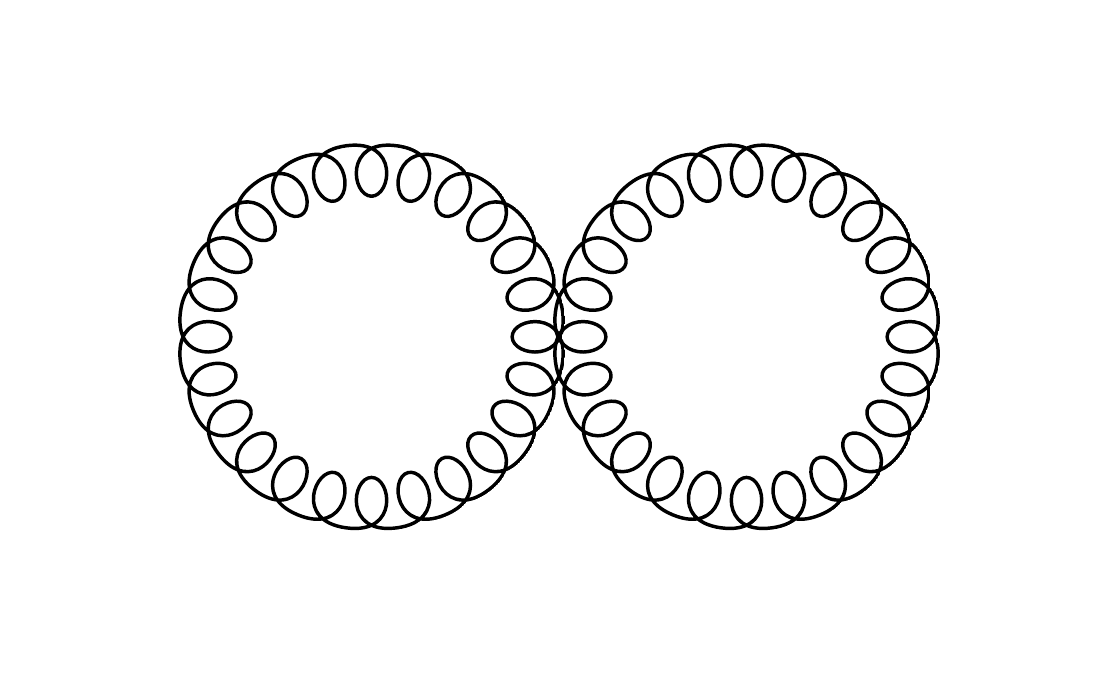}
			\includegraphics[width=2.5cm]{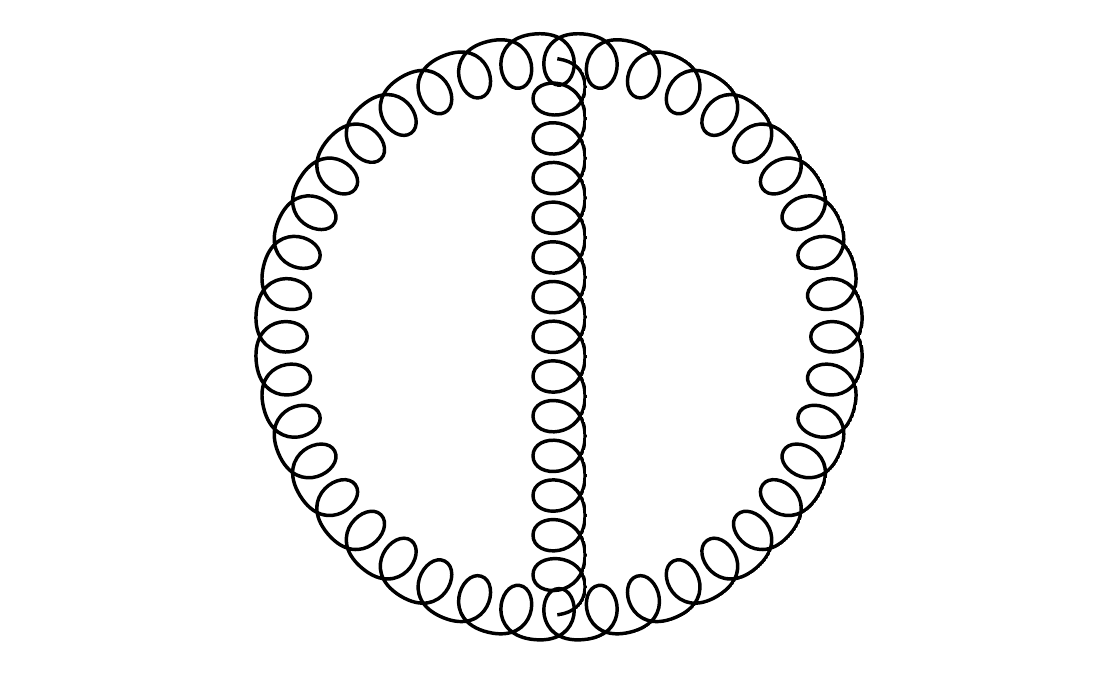}
			\includegraphics[width=2.5cm]{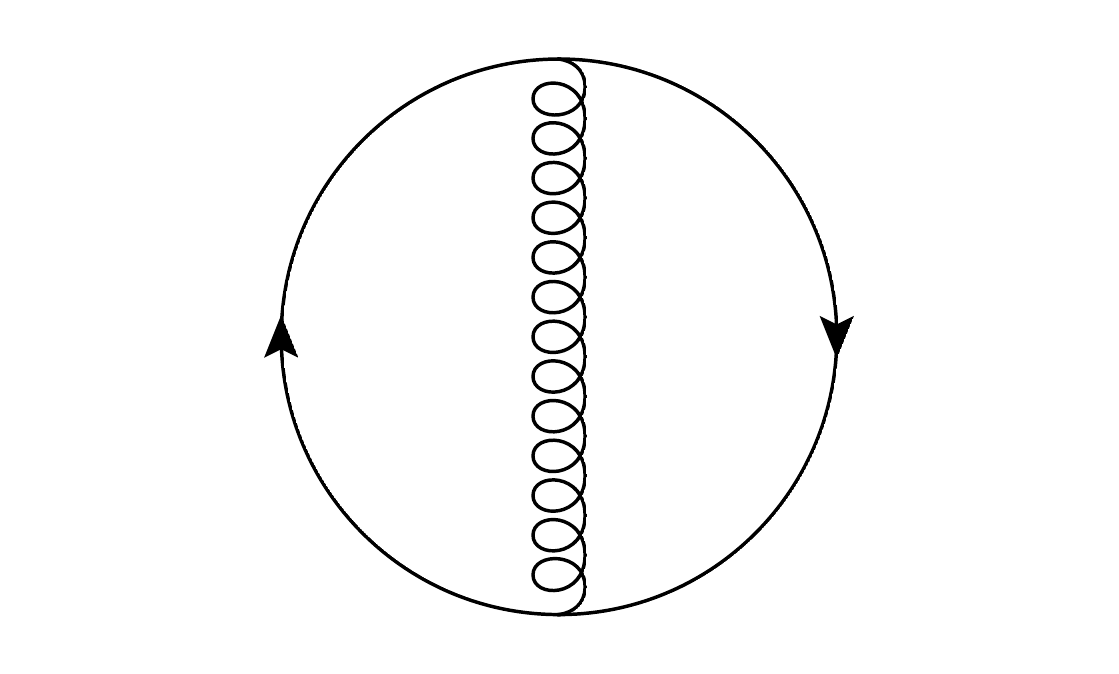}} &
			$\frac{\beta_{\alpha_g}^2}{\alpha_g}$ & $\beta_{\alpha_g}$ \\
		\hline
		\raisebox{-0.6cm}{\includegraphics[width=2.5cm]{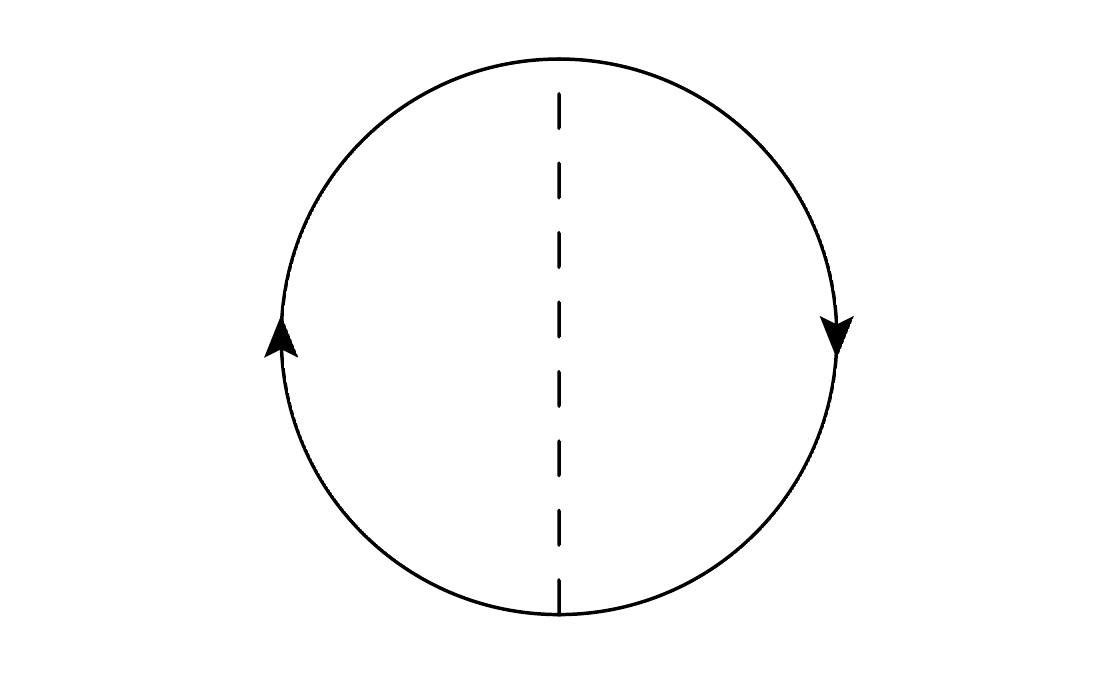}} &
			$\frac{\beta_{\alpha_y}^2}{\alpha_y}$ & $\beta_{\alpha_y}$ \\
		\hline\hline
		\raisebox{-0.6cm}{\includegraphics[width=2.5cm]{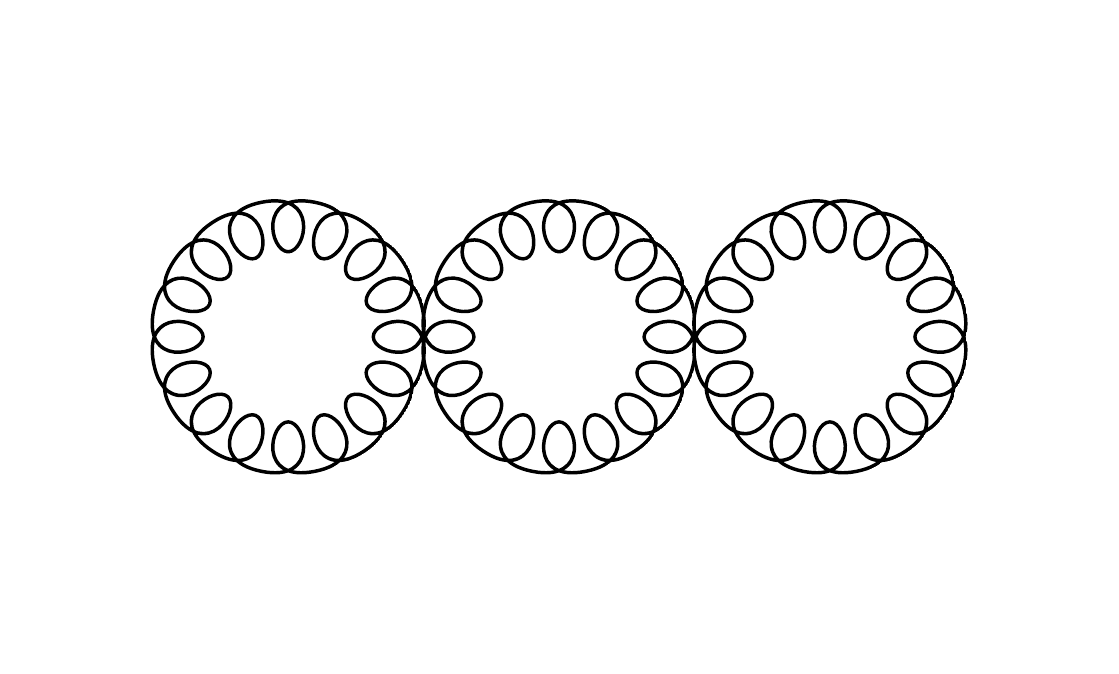}
			\includegraphics[width=2.5cm]{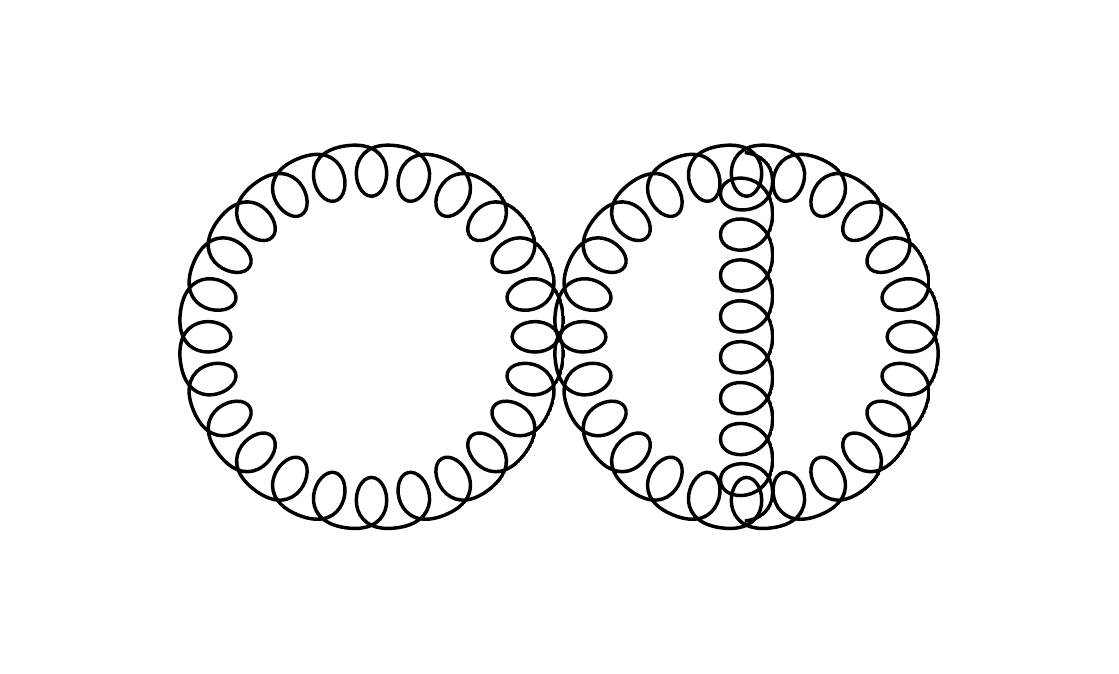}
			\includegraphics[width=2.5cm]{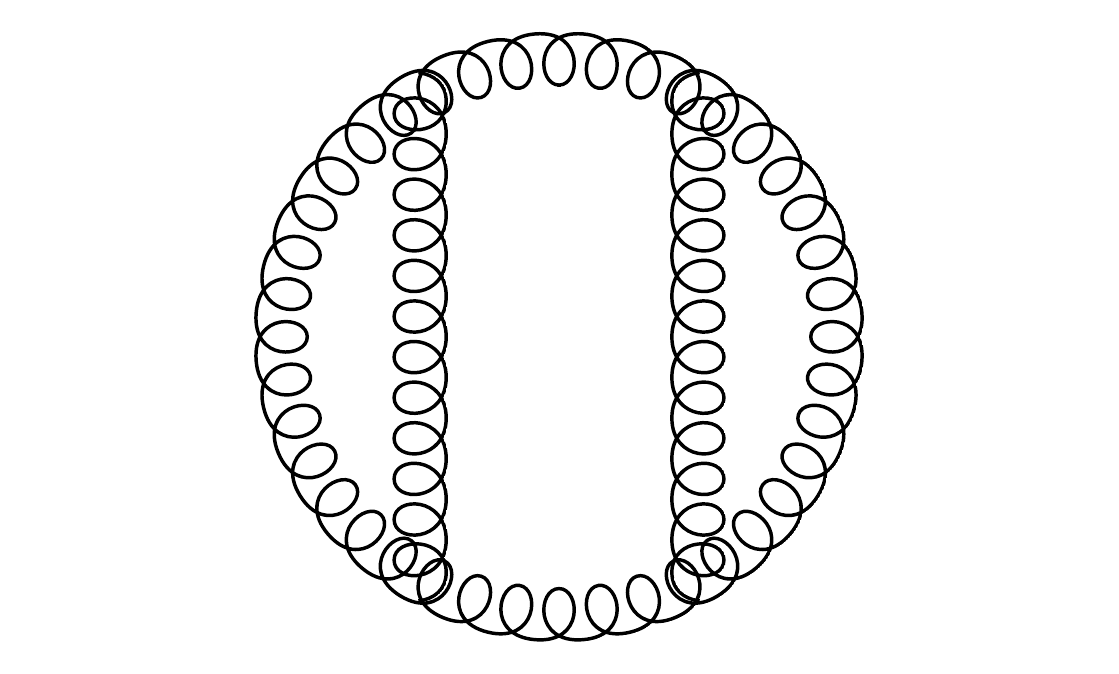}} && \\
		\raisebox{-0.6cm}{\includegraphics[width=2.5cm]{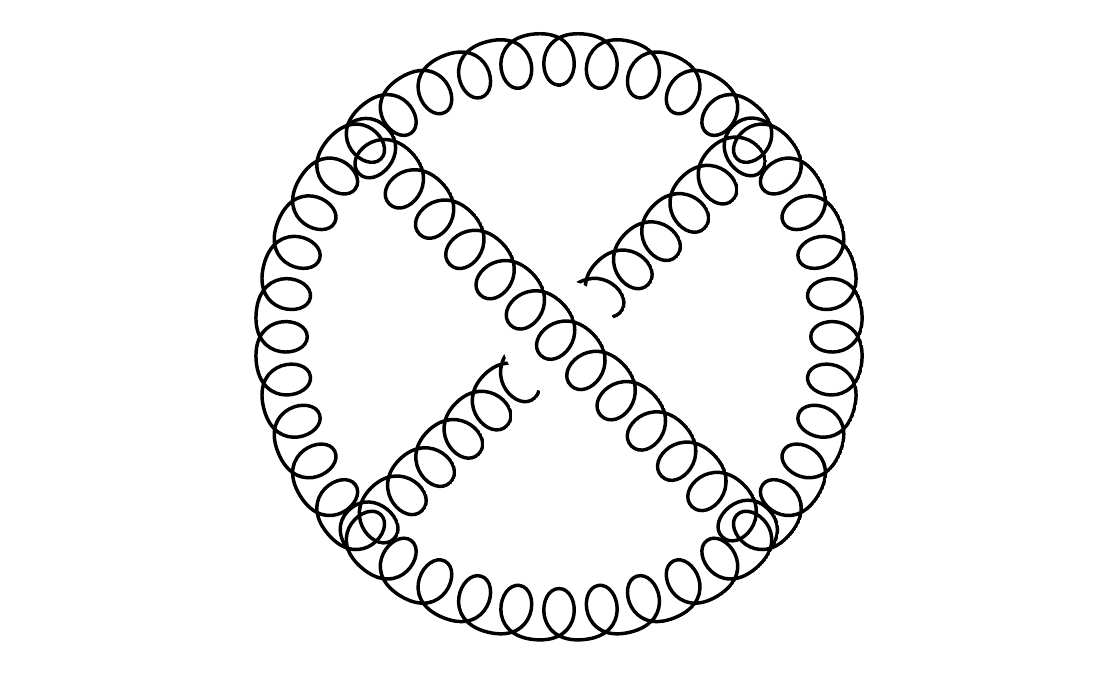}
			\includegraphics[width=2.5cm]{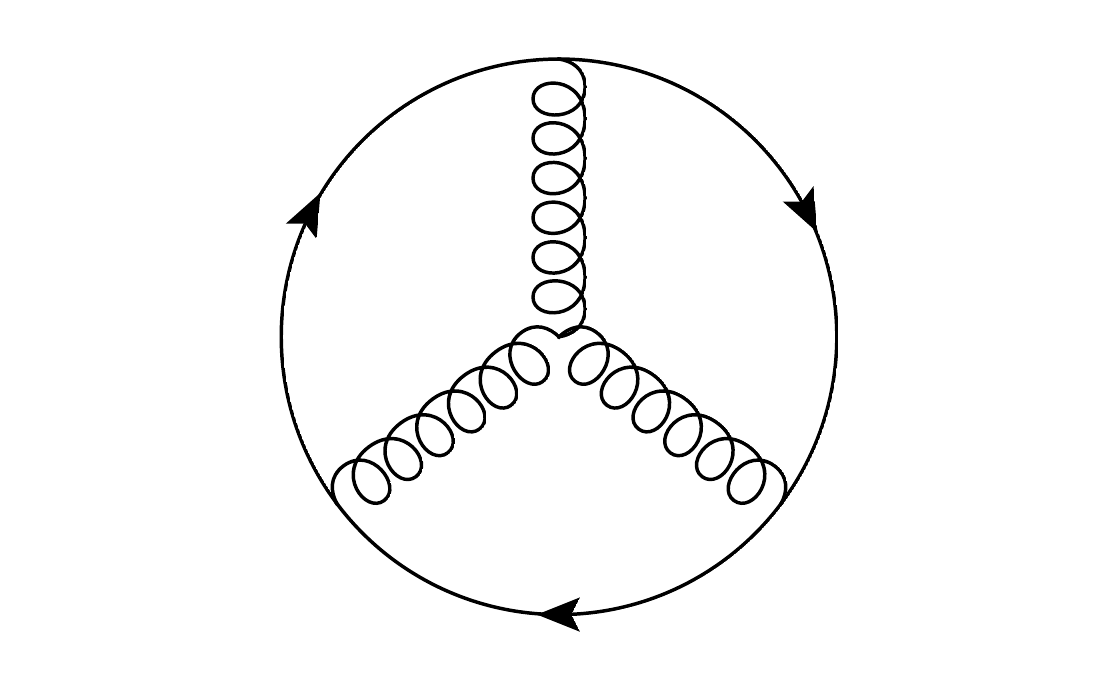}} \ldots &
			$\beta_{\alpha_g}^2$ &
			$\alpha_g \beta_{\alpha_g}$ \\
		\raisebox{-0.6cm}{\includegraphics[width=2.5cm]{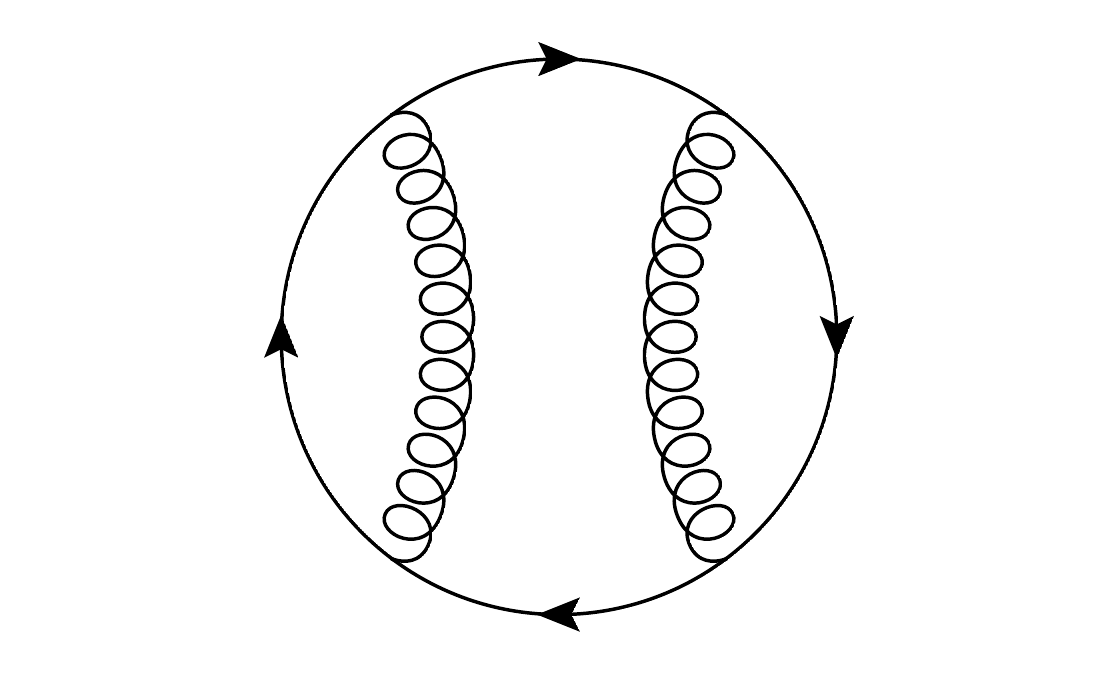}
			\includegraphics[width=2.5cm]{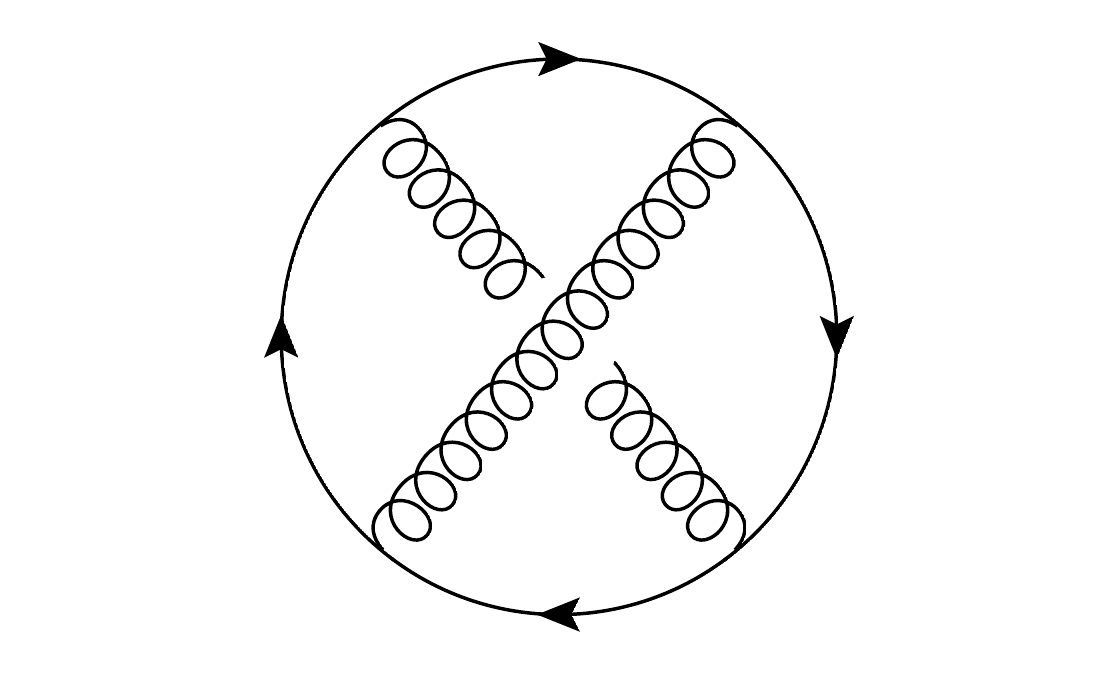}
			\includegraphics[width=2.5cm]{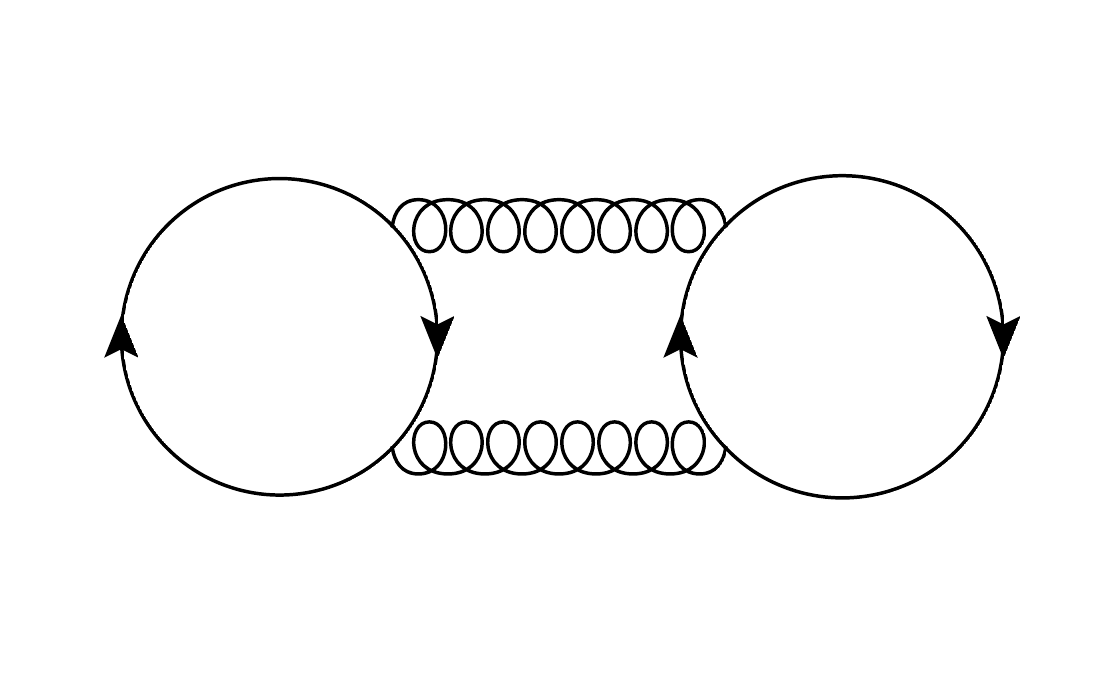}} && \\
		\hline
		\raisebox{-0.6cm}{\includegraphics[width=2.5cm]{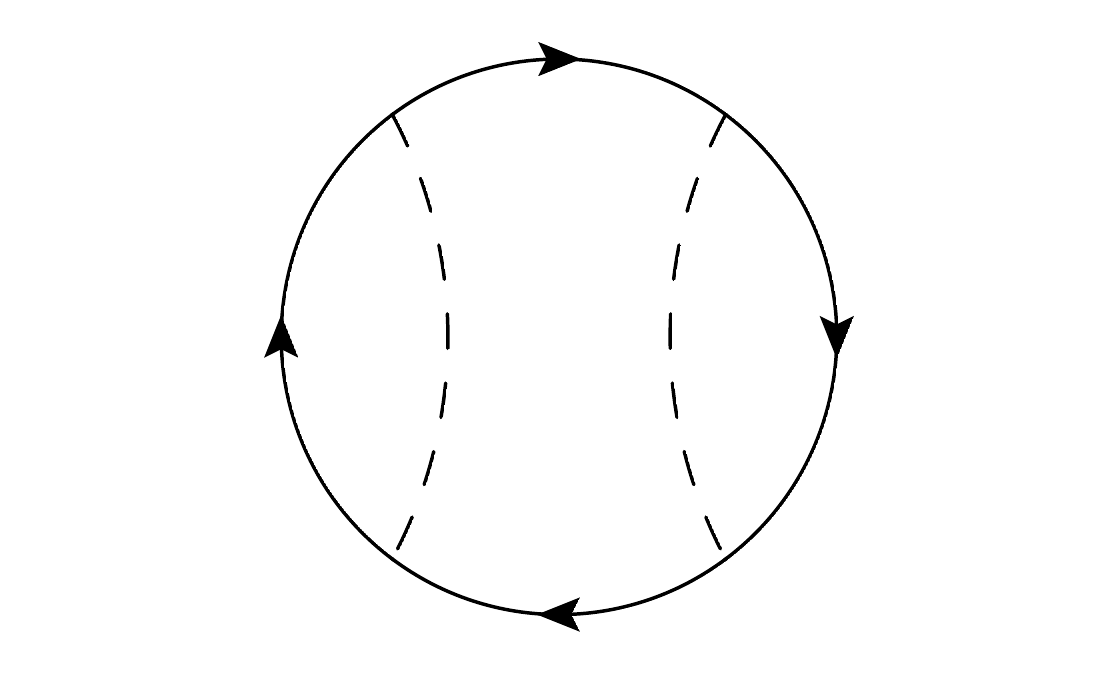}
			\includegraphics[width=2.5cm]{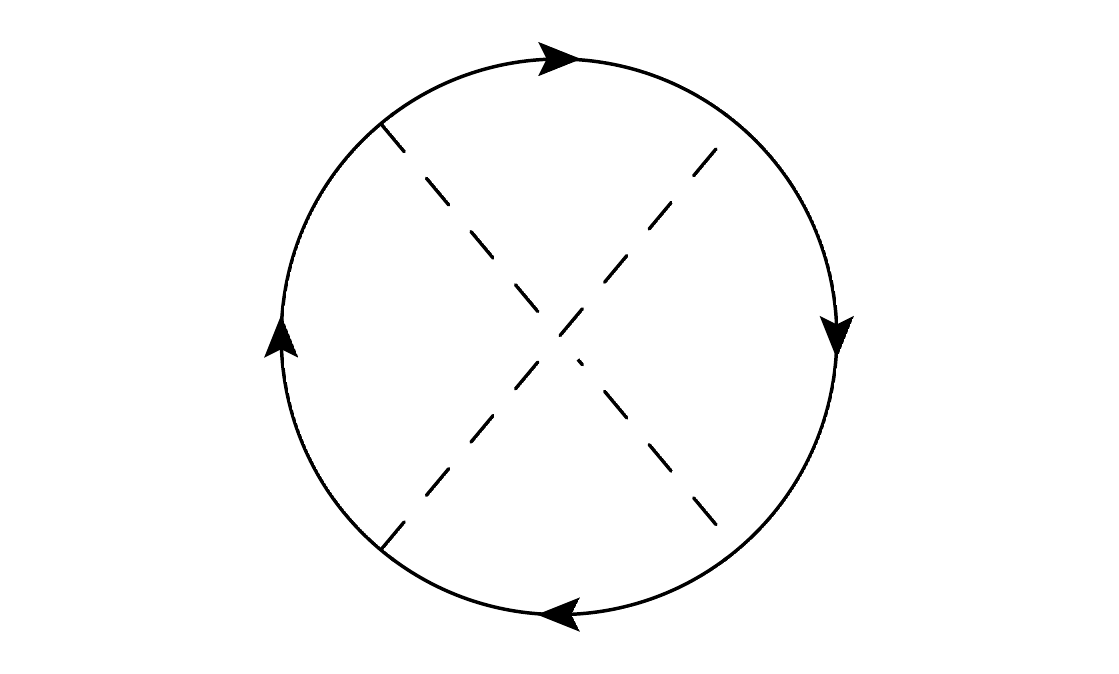}
			\includegraphics[width=2.5cm]{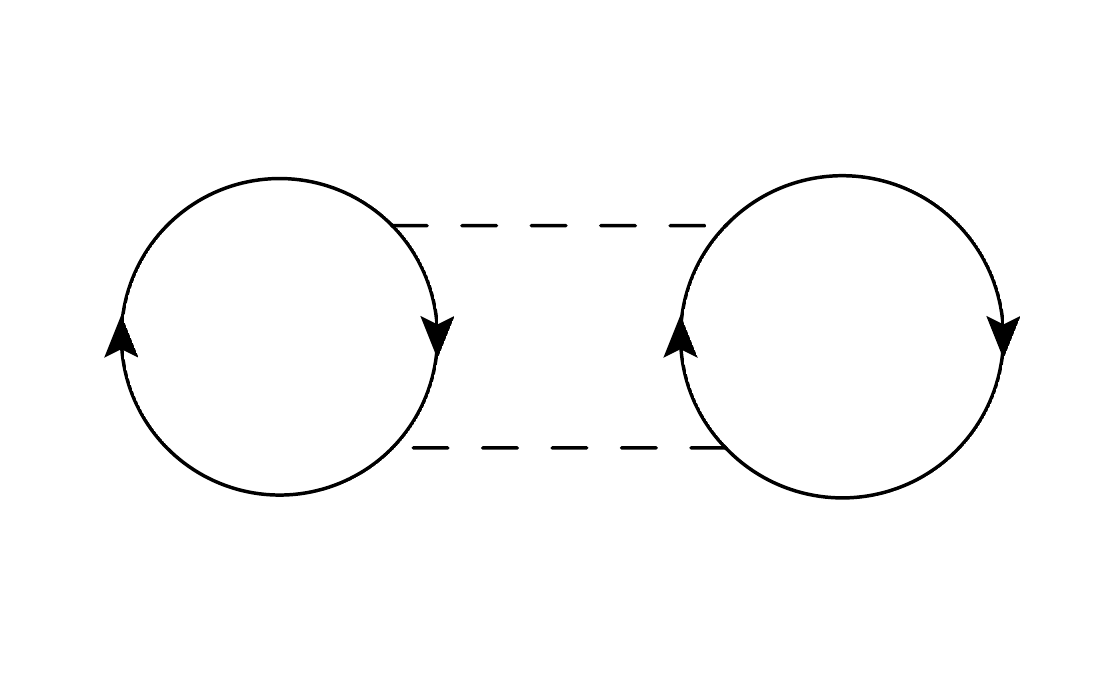}} &
			$\beta_{\alpha_y}^2$ &
			$\alpha_y \beta_{\alpha_y}$ \\
		\hline
		\raisebox{-0.6cm}{\includegraphics[width=2.5cm]{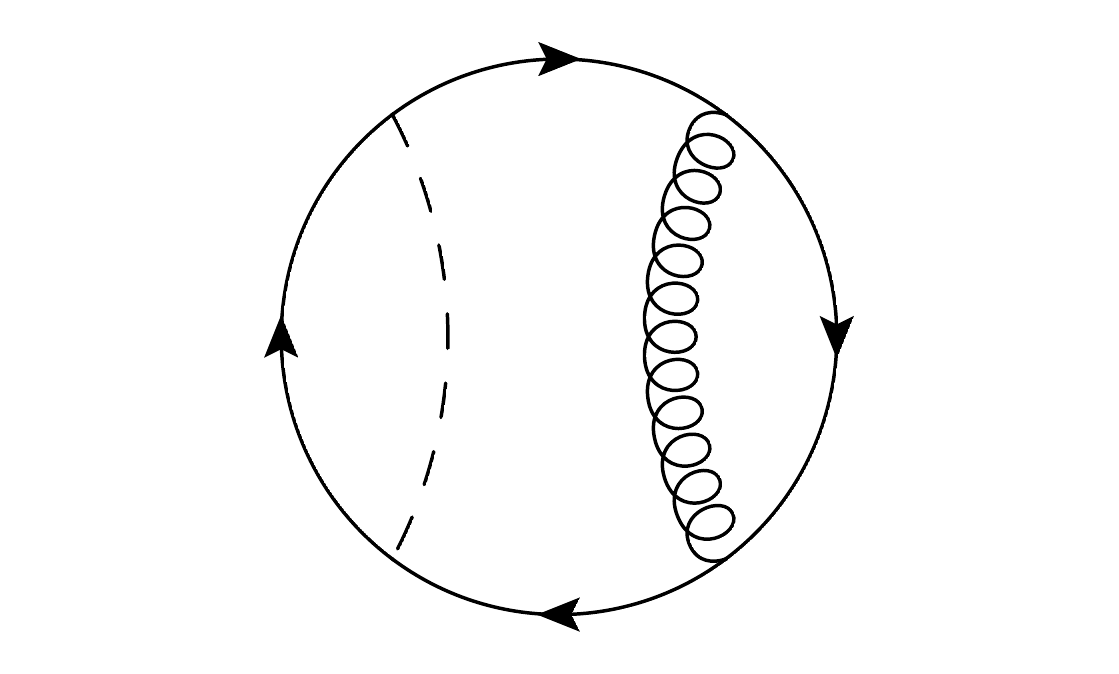}
			\includegraphics[width=2.5cm]{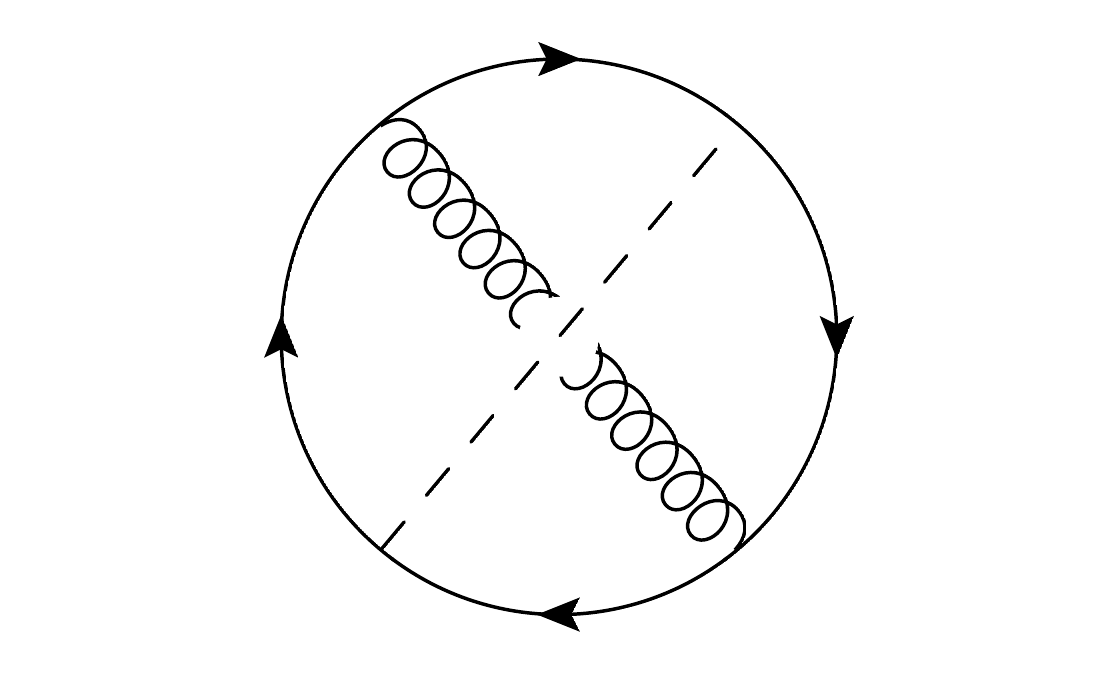}
			\includegraphics[width=2.5cm]{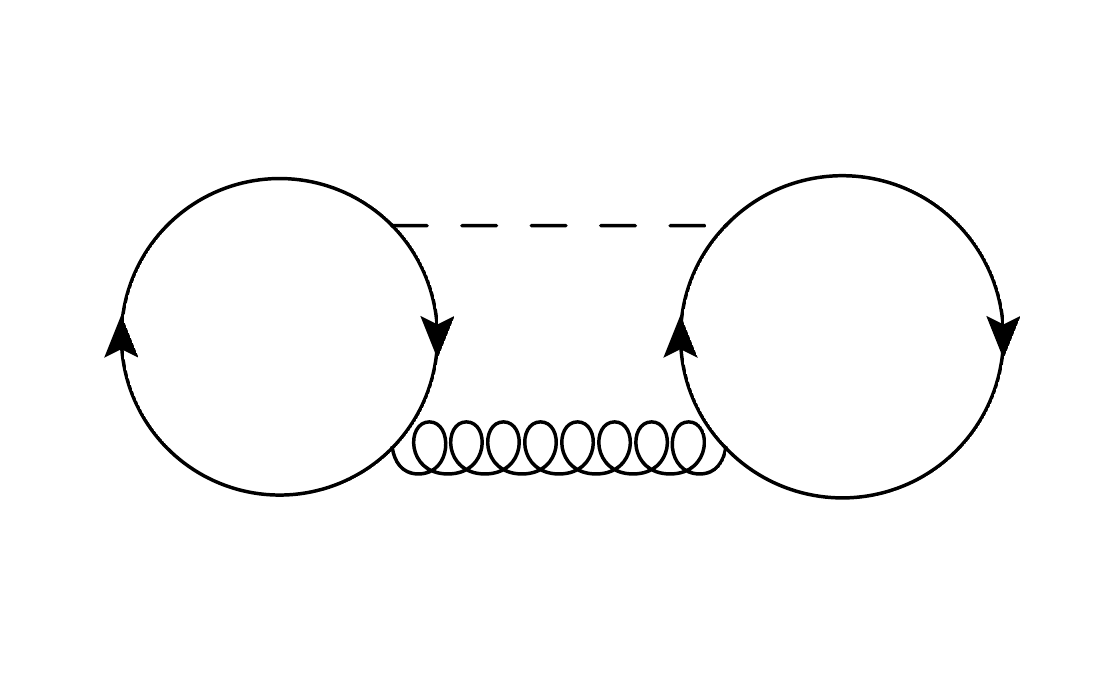}} &
			$\frac{\beta_{\alpha_g}^2}{\alpha_g} \alpha_y, ~~
				\frac{\beta_{\alpha_y}^2}{\alpha_y} \alpha_g, ~~
				\beta_{\alpha_g} \beta_{\alpha_y}$ &
			$\alpha_y \beta_{\alpha_g}, ~~ \alpha_g \beta_{\alpha_y}$ \\
		\hline
		\raisebox{-0.6cm}{\includegraphics[width=2.5cm]{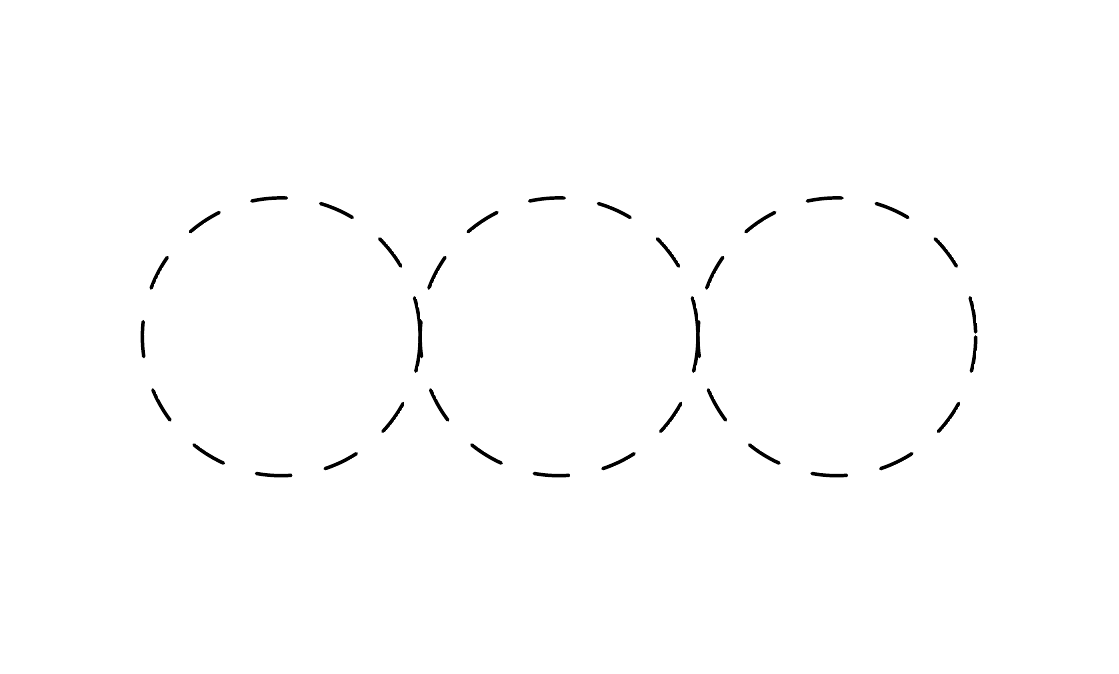}
			\includegraphics[width=2.5cm]{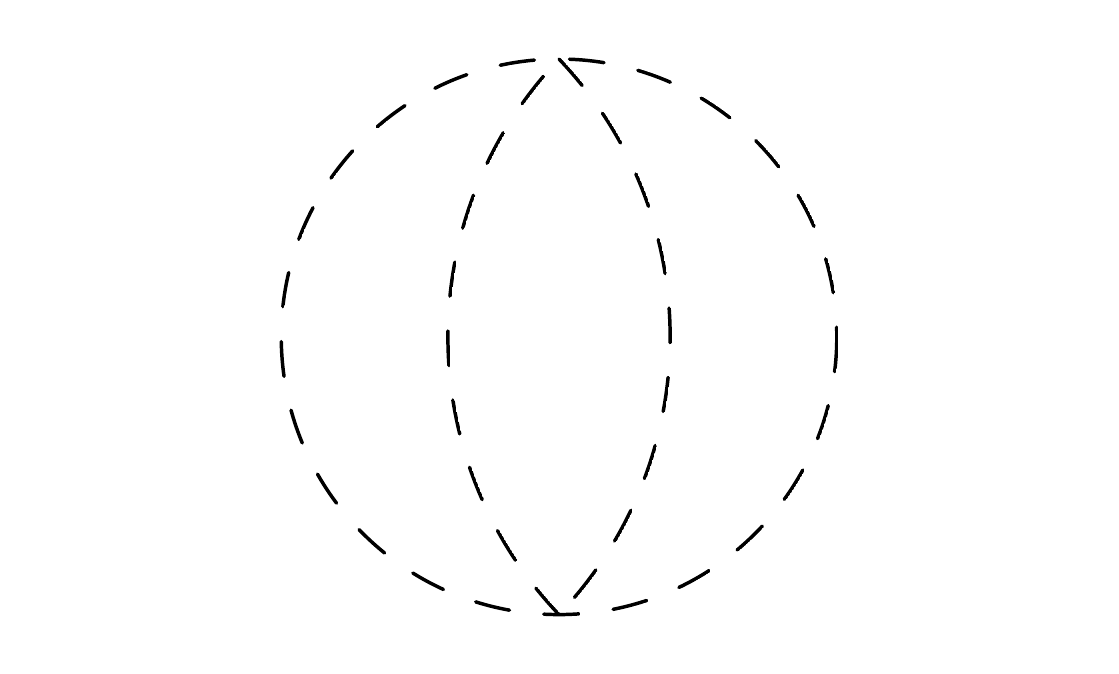}} &
			$\beta_{\alpha_\lambda}^2$ &
			$\alpha_\lambda \beta_{\alpha_\lambda}$ \\
		\hline
	\end{tabular}
	\caption{One, two and three--loop vacuum polarisation diagrams entering the computation of the metric (\ref{eq:metric}) and the one--form (\ref{eq:oneform}).}
	\label{tab:vacpoldiagrams}
\end{table}

\bibliography{biblio}

\end{document}